\documentclass[prd,aps,nofootinbib,nobibnotes,superscriptaddress,preprintnumbers,twocolumn]{revtex4}

\usepackage{color}
\usepackage{amsmath}
\usepackage{amsfonts}
\usepackage{graphicx}
\usepackage[dvipsnames]{xcolor}
\definecolor{refs}{RGB}{245,156,74}
\usepackage[colorlinks=true,hyperfootnotes=true,citecolor=cyan]{hyperref}

\usepackage{subcaption}
\usepackage{caption}
\usepackage{float}

\usepackage{physics}

\newcommand{\be}{\begin{equation}}
\newcommand{\ee}{\end{equation}}

\newcommand{\ricci}{\mathcal{R}}
\newcommand{\munu}[1]{{#1}_{\mu\nu}}
\newcommand{\cA}{\mathcal{A}}
\newcommand{\cH}{\mathcal{H}}

\newcommand{\tauexp}{\kappa}
\newcommand{\Rsun}{\hat{R}_\odot}
\newcommand{\qdot}{~.}
\newcommand{\qcom}{~,}
\newcommand{\qwith}{\quad\text{with}\quad}
\newcommand{\LCDM}{\Lambda\mathrm{CDM}}
\newcommand{\myeqref}{Eq.\eqref}
\newcommand{\myfigref}{Fig.\ref}

\newcommand{\cste}{\text{cste}}
\newcommand{\mysecref}{§\ref}
\newcommand{\rhoL}{\hat{\rho}_\Lambda}

\definecolor{tabblue}{HTML}{1f77b4}
\definecolor{taborange}{HTML}{ff7f0e}
\definecolor{tabgreen}{HTML}{2ca02c}
\definecolor{tabred}{HTML}{d62728}
\definecolor{tabpurple}{HTML}{9467bd}
\newcommand\crule[3][black]{\textcolor{#1}{\rule{#2}{#3}}}

\begin{document}

\title{Two-field Screening and its Cosmological Dynamics}

\author{Philippe Brax}
\affiliation{Institut de Physique Th\'eorique, Universit\'e Paris-Saclay,CEA, CNRS, F-91191 Gif-sur-Yvette Cedex, France.}
\affiliation{CERN, Theoretical Physics Department, Geneva, Switzerland.}
\author{Ayoub Ouazzani}
\affiliation{Institut de Physique Th\'eorique, Universit\'e Paris-Saclay,CEA, CNRS, F-91191 Gif-sur-Yvette Cedex, France.}
\affiliation{D\'epartement de Physique de l'Ecole Normale Supérieure, Paris.}

\begin{abstract}
    We consider the screening of the axio-dilaton fields when both the dilaton and the axion couple to matter with Yukawa couplings. We analyse the screening of the dilaton in the vicinity of a compact object  and find that this can only take place when  special boundary conditions at infinity are imposed. We study the cosmological dynamics of the axio-dilaton system when coupled to matter linearly and find that the special boundary conditions at infinity, which guarantee the screening of compact objects,  do not generically  emerge from cosmology. We analyse the background cosmology and the cosmological perturbations at late time in these models and show that  the Baryon Acoustic Oscillations constrain the coupling of the dilaton to matter to be  smaller than in its natural supergravity realisation. Moreover we find that the Hubble rate in the present Universe could deviate from the normalised Planck value, although by an amount too small to account for the $H_0$ tension, and that the growth of structure is  generically reduced compared to $\Lambda$CDM.
\end{abstract}

\maketitle
\section{Introduction}

General Relativity (GR) passes all the current gravitational tests in the solar system  with flying colours\cite{will2014confrontation}. Yet there are still  tensions on larger scales which may eventually necessitate to challenge   Einstein's theory of gravity. These are related to two classes of observations. First, there are long standing astrophysical results  that  imply the existence of dark matter \cite{Feng:2010gw}. Second, there is now robust evidence in favour of the acceleration of the expansion of the Universe \cite{Planck:2018vyg,Copeland:2006wr}. The latter  is best accounted for  in GR by adding a cosmological constant to the Lagrangian of the theory. However, understanding the origin of this constant energy density, e.g. from quantum field theoretic considerations, has proven to be fraught with difficulties \cite{Weinberg:1988cp}.
Modifying GR is not an easy task either \cite{deRham:2010kj}. We know from Lovelock's theorem \cite{Lovelock:1972vz} and,  in the effective field theory context, from Weinberg's theorem \cite{Weinberg:1965rz} that GR is unique in four dimensions provided  Lorentz invariance and the masslessness of the graviton hold \cite{forLovelock}.

One way to modify GR is to add additional fields. This is the route taken by  scalar-tensor theories  \cite{Damour:1992we} in which  one or more additional scalar fields are added to the setting and couple to matter. Such scalar fields appear in many proposed dynamical models attempting  to account for the late cosmic acceleration \cite{Brax_what_makes}. They also find theoretical motivation in the fact that they appear naturally in UV-complete theories such as string theory \cite{Damour:2002mi,Burgess_2022}. In particular, the swampland conjectures favour an explanation of the late-time acceleration of the Universe resulting from the non-trivial dynamics of string moduli \cite{Vafa:2005ui}. We will present the axio-dilaton which is one such scalar-tensor theories whose origin can be traced to a compactification of ten to four dimensions \cite{Burgess_2022}. 

Unfortunately, most scalar-tensor theories are  typically  ruled out by solar system tests of gravity. In their most naive form, consistency of scalar-tensor theories either requires them to have a small coupling to matter or the scalar must be stabilised with large masses leading to  short ranged interactions \cite{testing_screened}.

This can be illustrated with the Brans-Dicke theory (BD) and the action \cite{will2014confrontation,Avilez:2013dxa}
\begin{eqnarray}\label{eq:BD_lagrangian}
     \mathcal{S}_\mathrm{BD}=\int d^4x \sqrt{-g}M_p^2\left(\frac{\mathcal{R}}{2}-\frac{1}{2}\partial^\mu\varphi\partial_\mu\varphi-V(\varphi)\right) \nonumber \\
     + \mathcal{S}_m\left({g}^J_{\mu\nu},\,\psi_m\right) 
\end{eqnarray}
where $\psi_m$  represents the ordinary matter fields  and the Jordan frame metric is given by 
\begin{equation}
    {g}_{\mu\nu}^J=A^2(\varphi){g}_{\mu\nu} \qwith A(\varphi)=e^{\mathfrak{g}\varphi} 
\end{equation}
where $\mathfrak{g}$ is a coupling constant. The scalar couples to matter only via the metric ${g}_{\mu_\nu}^J$. The metric ${g}_{\mu\nu}$ is the Einstein metric for which the  Einstein equations take their usual form, i.e. matter  ``feels'' the geometry of the Jordan frame.

Constraints from solar system tests like those using measurements by the Cassini probe \cite{Cassini} can be understood using the Parameterized Post-Newtonian (PPN) formalism. It is enough here to define  two  important PPN parameters. We introduce the gravitational mass of the source $M$, and the PPN parameters $\gamma_{PPN}$, $\beta_{PPN}$ via the following parameterisation of the Jordan metric element in isotropic coordinates around a given compact object
\begin{eqnarray}
\label{eq:PPN}
&& ds_J^2=-\left(1-\frac{2GM}{r}+2\beta_{PPN}\left(\frac{GM}{r}\right)^2+{\cal O}\left(\frac{1}{r^3}\right)\right)d t^2\nonumber \\ 
&& +\left(1+2\gamma_{PPN}\frac{GM}{r}+{\cal O}\left(\frac{1}{r^2}\right)\right)\left( dr^2+r^2  d\Omega^2\right)\qdot\nonumber \\
\end{eqnarray}
In GR $\gamma_{PPN}=\beta_{PPN}=1$ whilst for the Brans-Dicke theory the deviation from GR is captured by $\gamma_{PPN}$, i.e.  $\beta_{PPN}=1$ and \cite{Burgess_2022}
\begin{equation}
    \gamma_{PPN}=\frac{1-2\mathfrak{g}^2}{1+2\mathfrak{g}^2}\qdot
\end{equation}
The Cassini probe on the other hand gives 
\cite{Cassini}
\begin{equation}\label{eq:Cassini}
\vert {\gamma_{PPN}-1}\vert <2.3\times 10^{-5}\qdot
\end{equation}
This constrains the coupling $\mathfrak{g}^2$ to be less than $10^{-5}$.

Larger deviations from GR can be reached in certain regimes when the theories are subject to  a \textit{screening mechanism}, i.e. a modification which allows the theory to evade solar system tests while having an order unity  coupling to matter whose role is important on large scales. There are now at least three types of screening mechanisms for single field models, i.e. the chameleon, K-mouflage and Vainshtein mechanisms \cite{Khoury:2003rn, Babichev:2009ee,Vainshtein:1972sx}. They all rely on higher derivatives and/or non-linearities of the kinetic terms and interacting potentials of the models \cite{testing_screened}.
In \cite{Burgess_2022}, a new screening mechanism was introduced, which relies on a second field that has a non-zero but small coupling to matter. The mechanism depends crucially on the interplay between the scalar profiles inside and outside matter \cite{Brax:2022vlf,Lacombe:2023qfx}. Here, we consider the situations where the two fields, the dilaton and the axion, have a linear coupling to matter that we denote by $\kappa$ and $\epsilon$. We focus on the small $\epsilon$ regime and vary $\kappa$ from small values up to unity which corresponds to its value in supergravity where
the dilaton plays the role of the volume modulus of string compactifications \cite{Cicoli:2023opf}. 
We find that for $\kappa=1$, screening of the dilaton around compact objects only takes place when the fields take particular values at infinity. These values should emerge from the cosmological dynamics. We then focus on the cosmology of these values where both $\kappa$ and the values of the coupling $\epsilon$ to both baryons and cold dark matter are varied. Generically, the field values in the present Universe do not satisfy the screening conditions. In fact, one must resort to yet unknown screening mechanisms for the axio-dilaton system in order to accommodate both local solar system tests and cosmological constraints such as the Baryon Acoustic Oscillations \cite{eBOSS:2020yzd}. 

The coupling of the axion to dark matter $\epsilon_C$ plays a crucial role cosmologically and can be of order unity.  On the other hand, we find that the coupling of the dilation $\kappa$ must be reduced locally to small values in order to pass solar system tests. In this paper, we will consider two likely scenarios. The first one is that the coupling $\kappa$ is determined locally to be extremely small, of the order of $10^{-3}$, and that most of the cosmological dynamics are due to the coupling of the axion $\epsilon_C$ in a manner reminiscent of coupled quintessence \cite{Amendola:1999er} although in a multi-field setting here \cite{Amendola:2014kwa}. This could be achieved if another field $\chi$ drove the coupling $\kappa$ to such values dynamically in the whole Universe. Another possibility could be that the coupling $\kappa$ is small locally but allowed to take larger values cosmologically. This could happen if the field $\chi$ only made $\kappa$ small locally. In this second scenario, we find that the BAO constraints on late time cosmology are so stringent that $\kappa$ cannot be taken of order unity as in the original supergravity model. In this setting, we consider the allowed deviations of $H_0$ from their  values as calibrated by the Planck satellite experiment and find that only a few percent of discrepancy are allowed. This is much less than the current $H_0$ tension \cite{Verde:2019ivm,hubble_tension}. Finally, we notice that the linear growth in these models is reduced compared to the $\Lambda$CDM case, despite the existence of  attractive fifth forces due to the dilaton and the axion. This could provide a solution to the $\sigma_8$ tension \cite{Barros:2018efl} where the observed amount of clustering is reduced compared to the expected one from early times \cite{DiValentino:2020vvd,Nunes:2021ipq}. The precise study of this possibility is left to future work. 

The scenario that we present here enlarges the usual single-field class of models for late time cosmology. Such multi-field generalisations could prove useful in view of future measurement and present cosmological tensions \cite{Amendola:2016saw}.

The paper is arranged as follows. In section \ref{sec:AxioDilatonTh}, we present the axio-dilaton model. In section  \ref{sec:sourceandscreening}, we consider the screening of a compact objects. We then discuss the cosmology of these models in section \ref{sec:bcosmology}.

\section{The axio-dilaton theory}\label{sec:AxioDilatonTh}
\subsection{The Lagrangian}
The axio-dilaton theory contains two scalar fields, the dilaton $\tau>0$ and the axion $a$. The dilaton $\tau$ couples to matter only through the Jordan frame metric while $a$ is directly coupled to matter. The difference will be made clear below where we will construct an effective metric which will mediate the coupling of both scalars to matter.  The action of the theory is
\begin{equation}\label{eq:Lagrangian}
\mathcal{S}=\int d^4x \sqrt{-g}M_p^2\left[\frac{\mathcal{R}}{2}-\frac{3}{4}\left(\frac{\partial^\mu\tau\partial_\mu\tau+\partial^\mu a \partial_\mu a}{\tau^2}\right)\right]+\mathcal{S}_m \qcom
\end{equation}
with
\begin{eqnarray}
&&\mathcal{S}_m=\mathcal{S}_m({g}_{\mu\nu}^J,\,a,\,\psi_m),\qquad {g}_{\mu_\nu}^J=A(\tau)^2{g}_{\mu\nu}  \nonumber\\ &&A(\tau)=\tau^{-\tauexp/2}. \nonumber \\
\end{eqnarray}
 $\tauexp$ is a coupling constant. The theory studied in \cite{Burgess_2022} corresponds to $\tauexp=1$ and is associated to a supergravity model of string theory origin with a K\"ahler potential $K=-3 \ln (\mathcal{T}+\overline{\mathcal{T}})$ and a coupling to matter determined by $A=e^{K/6}$. In this setting, the dilaton can be seen as the volume modulus of a 6d compatification of string theory from 10d to 4d.  Introducing $\tauexp$ enables one to tune the matter-dilaton coupling and make the  solar system tests of gravity easier to satisfy. We will see that screening is compulsory for the model to pass the solar system tests of gravitation.

The two fields can be viewed as the real and imaginary parts of a complex field $\mathcal{T}=\frac{1}{2}(\tau+i a)$ whose Lagrangian is
\begin{equation}
\mathcal{L}=\sqrt{-g}M_p^2\left(\frac{\mathcal{R}}{2}-\frac{3\partial^\mu\mathcal{T}\partial_\mu\overline{\mathcal{T}}}{(\mathcal{T}+\overline{\mathcal{T}})^2}\right)+\mathcal{L}_m \qdot
\end{equation}
In this model, 
we define the usual stress-energy tensor and the coupling of $a$ to matter
\begin{equation}
T^{\mu\nu}\equiv -\frac{2}{\sqrt{-g}}\frac{\delta S_m}{\delta g_{\mu\nu}}, \  \mathcal{A}\equiv -\frac{2}{\sqrt{-g}}\frac{\delta S_m}{\delta a} \qdot
\end{equation}
We will use for the trace the notation $T\equiv g_{\mu\nu} T^{\mu\nu}$. Notice that the matter action depends on the axion field in a non-trivial manner and not via the Jordan metric. This will have drastic consequences that we will unravel below. 

There are two Klein-Gordon equations for the two scalar fields, i.e. 
\begin{equation}\label{eq:KGtau}
\Box \tau -\frac{1}{\tau}(\partial_\mu\tau\partial^\mu\tau-\partial_\mu a\partial^\mu a)-\tauexp\frac{\tau}{3 M_p^2}T=0 \qcom
\end{equation}
and 
\begin{equation}\label{eq:KGa}
\Box a -\frac{2}{\tau}\partial_\mu\tau\partial^\mu a+\frac{\tau^2}{3 M_p^2}\cA=0 \qdot
\end{equation}
It is convenient to  introduce the dilaton field $\varphi$ such that $\tau=e^\varphi$ leading to
\begin{equation}\label{eq:KGphi}
\Box \varphi+(\partial_\mu a\partial^\mu a)e^{-2\varphi}-\frac{\tauexp}{3 M_p^2}T=0\qdot
\end{equation}
The Einstein equation is simply
\begin{equation}\label{eq:Einstein}
\ricci_{\mu\nu}-\frac{3}{2\tau^2}(\partial_\mu\tau\partial_\nu\tau+\partial_\mu a\partial_\nu a)-\frac{1}{M_p^2}(T_{\mu\nu}-\frac{1}{2}T{g}_{\mu\nu})=0 \qdot
\end{equation}
where we have separated the matter energy momentum tensor from the scalar one.

\subsection{The symmetries}
In the absence of matter, the theory is invariant under a $SL(2,\mathbb{R})$ group
whose origin can be traced back to supergravity. Indeed, the kinetic term of the fields is invariant under 
\begin{equation}
    \mathcal{T}\longrightarrow\frac{a\mathcal{T}-ib}{ic\mathcal{T}+d}\qquad\text{provided}\qquad ad-bd=1
\end{equation}
corresponding to a K\"ahler transformation of the theory.
There are thus three conserved currents, corresponding to the dimension of the symmetry group in the absence of matter. We can choose as a basis for these currents
\begin{itemize}
    \item The axion shift symmetry $\mathcal{T}\rightarrow\mathcal{T}-ib$ ($a=c=0$, $d=1$):
    \begin{equation}
        J^\mu_A=\frac{\partial^\mu a}{\tau^2}\qdot
    \end{equation}
    \item The rescaling symmetry $\mathcal{T}\rightarrow a\mathcal{T}$ ($b=c=0$, $d=1$):
    \begin{equation}
        J^\mu_S=\frac{\partial^\mu \tau}{\tau}+\frac{a\partial^\mu a}{\tau^2}\qdot
    \end{equation}
    \item The non-linear symmetry $\mathcal{T}\rightarrow\mathcal{T}-ic\mathcal{T}^2$ ($a=d=1$, $b=0$, $c\ll 1$):
    \begin{equation}
        J^\mu_N=\frac{\tau^2-a^2}{\tau^2}\partial^\mu a-2a\frac{\partial^\mu \tau}{\tau}\qdot
    \end{equation}
\end{itemize}
From the Klein-Gordon equations we can directly obtain the (non-)conservation laws
\begin{eqnarray}\label{eq:currents}
    &&\nabla_\mu J^\mu_A=-\frac{\cA}{3M_p^2}\qcom\qquad\nabla_\mu J^\mu_S=\frac{\tauexp T-a{\cal A}}{3M_p^2}\qcom \nonumber \\ &&\nabla_\mu J^\mu_N=\frac{(a^2-\tau^2)\cA-2a\tauexp T}{3M_p^2}\qdot\nonumber \\
\end{eqnarray}
As can be seen, matter breaks the whole symmetry group as none of the three currents is conserved anymore.

When $\mathcal{S}_m$ does not depend on $a$, i.e. $\cA=0$, the axio-dilation theory is equivalent to a Brans-Dicke theory. Indeed when $\cA=0$ we have the axion solution $a=\cste$, and so \myeqref{eq:KGphi} becomes
\begin{equation}
    \Box\varphi-\frac{\tauexp}{3 M_p^2}T=0\qdot
\end{equation}
Similarly, the BD Lagrangian \eqref{eq:afromtau} gives the Klein-Gordon equation 
\begin{equation}
    \Box\varphi_\mathrm{BD}+\frac{\mathfrak{g}}{M_p^2}T=0\qdot
\end{equation}
 Matching the two Weyl factors $e^{-\tauexp\varphi/2}=e^{\mathfrak{g}\varphi_\mathrm{BD}}$ we get $\varphi_\mathrm{BD}=(-\tauexp/2\mathfrak{g})\varphi$. Combining the two Klein-Gordon equations
\begin{equation}
    \left(\frac{-\tauexp}{2\mathfrak{g}}+\frac{3\mathfrak{g}}{\tauexp}\right)\Box\varphi=0 \implies \mathfrak{g}^2=\tauexp^2/6\qdot
\end{equation}
When $\kappa=1$, we find that the coupling reduces to $1/\sqrt 6$, i.e. the same as for $f(R)$ and massive gravity.

\section{Non-relativistic source and screening}\label{sec:sourceandscreening}

Screening requires to study the gravitational physics around objects like the Sun. We model the Sun and other compact objects such as the Earth or the Moon as  non-relativistic sources, i.e the only non-zero component of their stress-energy tensor is $T_{00}\equiv \rho$. We also assume that  they are static and spherically symmetric. We look for space-time solutions with the same symmetries that we chart with isotropic coordinates
\begin{equation}\label{eq:metric_iso}
g_{\mu \nu} d x^\mu d x^\nu = -e^{2u(r)}d t^2+e^{2w(r)}\left(d r^2+r^2d \Omega^2\right) \qdot
\end{equation}
The Jordan metric is obtained by multiplying this line element by the coupling function $A^2$.
\subsection{Exterior solution}\label{sec:ext_solution}
We will simplify the setting  by considering non-relativistic objects with a small Newtonian potential. As a result, we will approximate the Klein-Gordon equations using a flat background $\boldsymbol{g}=\boldsymbol{\eta}$ where $\eta_{\mu\nu}$ is the Minkowski metric tensor. The validity of the approximation is evaluated a posteriori.

In the following,  primes will refer to derivatives with respect to $r$. We assume spherical symmetry. The resulting Klein-Gordon equations are  
\begin{equation}\label{eq:flatKGtau}
\tau''+\frac{2\tau'}{r}-\frac{\tau'^2-{a'}^2}{\tau}+\tauexp\frac{\tau\rho}{3M_p^2}=0 \qcom
\end{equation}
and 
\begin{equation}\label{eq:flatKGa}
a''+\frac{2a'}{r}-\frac{2\tau'a'}{\tau}+\frac{\tau^2\cA}{3M_p^2}=0 \qdot
\end{equation}
Here $\rho$ is the matter density inside or outside the object.
Instead of using these equations directly, we can use the equations for the currents \eqref{eq:currents}. In the absence of matter, i.e. outside the objects,  the currents are conserved
\begin{align}
    r^2\frac{a'}{\tau^2}=C_A\qcom\label{eq:C_A}\\
    r^2\left(\frac{\tau'}{\tau}+\frac{aa'}{\tau^2}\right)=C_S\qcom\\
    r^2\left(\frac{(\tau^2-a^2)a'}{\tau^2}-\frac{2a\tau'}{\tau}\right)=C_N\qdot
\end{align}
These constants are fixed by the conditions inside the source.

We are interested in the case where $a\neq\cste$. So $C_A\neq 0$. Defining $\gamma\equiv C_A$, $\alpha \equiv C_S/C_A$, $\beta\equiv (C_S/C_A)^2+C_N/C_A$, we obtain
\begin{equation}
    \tau^2+(a-\alpha)^2=\beta^2\qdot
\end{equation}
where $\tau$ and $a$ thus evolve on a circle in the $\tau-a$ plane. We can thus  eliminate $\tau$ in \myeqref{eq:C_A} and get
\begin{equation}
    a'=\frac{\gamma}{r^2}(\beta^2-(a-\alpha)^2)\qdot
\end{equation}
We integrate this to obtain the axion profile
\begin{equation}\label{eq:axion_ext}
    a=\alpha-\beta\tanh{X(r)} \qwith X(r)=\frac{\gamma\beta}{r}+\delta\qcom
\end{equation}
where $\delta$ is a new integration constant. Using again \myeqref{eq:C_A} we finally get
\begin{equation}\label{eq:tau_ext}
    \tau=\frac{\beta}{\cosh{X(r)}}\qdot
\end{equation}
Some of the integration constants are fixed by the boundary conditions inside the source. Indeed, from the currents \eqref{eq:currents}, we see that
\begin{equation}\label{eq:gamma_as_int}
    \gamma=C_A=-\frac{1}{3M_p^2}\int_0^R dr r^2\cA(r)\qcom
\end{equation}
\begin{equation}\label{eq:gammaalpha_as_int}
    \gamma\alpha=C_S=-\frac{1}{3M_p^2}\int_0^R d r r^2(\tauexp{}\rho(r)+a(r)\cA(r))\qdot
\end{equation}
On the other hand, $\beta$ and $\delta$ can only be fixed by  the values  of the fields at infinity
\begin{equation}\label{eq:BQ_infty}
a_\infty=\alpha-\beta\tanh{\delta}, \ \beta=\tau_\infty\cosh{\delta}=\sqrt{\tau_\infty^2+(\alpha-a_\infty)^2}\qdot
\end{equation}
As shown in appendix \ref{app}, these solutions in the  flat background approximation are valid as long as 
\begin{equation}\label{eq:flatapprox_condi}
r\gg GM \qquad \text{and} \qquad r\gg\vert\gamma\beta\vert \qdot
\end{equation}
The first condition corresponds to being far from the Schwarzschild radius.

\subsection{Screening}\label{sec:screening}

We first investigate screening in the Jordan frame where we will extract the PPN parameters from the Jordan metric
\begin{equation}
\begin{aligned}
&{g}_{\mu\nu}^J dx^\mu dx^\nu =A^2{g}_{\mu\nu}dx^\mu dx^\nu\\
&=-\left(1-\frac{2GM_g}{r}+2\beta_{PPN}\left(\frac{GM_g}{r}\right)^2+{\cal{O}}\left(\frac{1}{r^3}\right)\right)dt^2\\
&+\left(1+2\gamma_{PPN}\frac{GM_g}{r}+{\cal{O}}\left(\frac{1}{r^2}\right)\right)(dr^2+r^2 d\Omega^2)\qdot
\end{aligned}
\end{equation}
where $A^2=1/\tau^{\tauexp}$ and $g_{\mu\nu}$ is given by \eqref{eq:metric_iso}. $M_g$ is the gravitational mass as defined in the PPN formalism. We have that $M_g\neq M=\int_\text{source}\rho$, i.e. the mass of the object in the Jordan frame is renormalised by the presence of the scalar fields. 

In the absence of fields, the Einstein frame metric $g_{\mu\nu}$ is the Schwarzschild metric. In the presence of the fields the  Einstein equations are modified and we expand
\begin{align}\label{eq:schwarz}
e^{2u}&=1-\frac{2 l}{r}+\frac{2 l^2}{r^2}+{\cal O}\left(\frac{1}{r^3}\right) \qcom\\
e^{2w}&=1+\frac{2 l}{r}+{\cal O}\left(\frac{1}{r^3}\right)\qdot 
\end{align}
where $l=GM$. Expanding the conformal factor $A^2$ in inverse powers of the distance, we have
\begin{equation}
    A^2=A_\infty^2\left(1-\alpha_1/{r}+{\alpha_2}/{r^2}+{\cal O}\left(\frac{1}{r^3}\right)\right)\qcom
\end{equation}
from which we deduce that
\begin{align}
\gamma_{PPN}&=\frac{1-\frac{\alpha_1}{2l}}{1+\frac{\alpha_1}{2l}}\qcom\\
\beta_{PPN}&=\frac{l^2+l\alpha_1+\frac{1}{2}\alpha_2}{(l+\frac{1}{2}\alpha_1)^2}\qdot
\end{align}
For the axio-dilaton theory the conformal factor is given by
\begin{equation}
A^2=\tau^{-\tauexp}=\left(\frac{\cosh(\frac{\beta\gamma}{r}+\delta)}{\beta}\right)^\tauexp 
\end{equation}
where we have used the explicit solution for $\tau (r)$. 
Expanding in $1/r$ we find the coefficients 
\begin{eqnarray}
&&\alpha_1=-\tauexp\beta\gamma\tanh{\delta} \qquad, \nonumber \\ &&\alpha_2=\frac{\tauexp\beta^2\gamma^2}{2}((\tauexp-1)\tanh^2{\delta}+1)\qdot
\end{eqnarray}
In the following, we will always choose the following ansatz following \cite{Burgess_2022}
\begin{equation}
    \cA=-\epsilon T\qcom
\end{equation}
where $\epsilon$ is a small constant and $T$ the trace of the energy momentum tensor. In the case of static sources this reduces to $\cA= \epsilon \rho$. 
As we shall see, this choice is not innocent as this will bring back the two-field model within the realm of the scalar-tensor theories with an effective coupling to both the dilaton and the axion. We will make this explicit in the following. 
From \myeqref{eq:gamma_as_int} we obtain 
\begin{equation}
\gamma=-\frac{2\epsilon GM}{3}=-\frac{2\epsilon l}{3}\qdot
\end{equation}
and in the Jordan frame the PPN parameters become
\begin{equation}\label{eq:gamma_PPN}
\gamma_{PPN}=\frac{3-\epsilon\tauexp\beta\tanh{\delta}}{3+\epsilon\tauexp\beta\tanh{\delta}}\qcom
\end{equation}
and 
\begin{equation}\label{eq:beta_PPN}
\beta_{PPN}=1+\frac{\tauexp\epsilon^2\beta^2(1-\tanh^2{\delta})}{(3+\tauexp\epsilon\beta\tanh{\delta})^2}\qdot\nonumber \\
\end{equation}
Hence by taking $\epsilon$ small GR is recovered as long as $\beta$ does not increase accordingly. This is the essence of this new type of screening which corresponds to a non-uniform limit $\epsilon\to 0$. Indeed, when $\epsilon=0$ the model is equivalent to a Brans-Dicke model with a coupling $\kappa/\sqrt 6$ which needs to be small enough to satisfy the solar system tests of gravity. But it turns out that the non-uniform limit $\epsilon \to 0$ requires a particular choice of the boundary conditions at infinity which are not generic. We will come back to this point in the next section.

\subsection{The interior solution}
We now  solve \myeqref{eq:flatKGtau} and \myeqref{eq:flatKGa} inside the source of radius $R$, with the boundary  conditions $\tau'(0)=a'(0)=0$ at the origin. We assume a uniform density inside the body as a simplifying assumption and  a coupling $\cA$ with the same profile as $\rho$
\begin{equation}
  \left\{
    \begin{aligned}
      \rho & \equiv \rho_0 = \frac{M}{\frac{4}{3}\pi R^3} \\
      \cA & = \epsilon \rho_0
    \end{aligned}
  \right.  \qdot
\end{equation}
The dynamical equations cannot be solved exactly. We will  get perturbative  solutions in $\epsilon$ as we have seen that a small coupling $\epsilon$ is required to screen in the Jordan frame. 

\subsubsection{The dilaton dynamics}
The equation for the current $J_A^\mu$ from \myeqref{eq:currents} gives 
\begin{equation}
  \left(r^2\frac{a'}{\tau^2}\right)'=-\frac{r^2\cA}{3M_p^2} \qdot 
\end{equation}
Assuming  regularity for the fields at the origin, we obtain
\begin{equation}
 \frac{a'(r)}{\tau^2(r)}=-\frac{\epsilon\rho_0}{3M_p^2r^2}\frac{r^3}{3} \qdot 
\end{equation}
We define $m^2=\rho_0/3M_p^2$ leading to 
\begin{equation}\label{eq:afromtau}
a'=-\frac{\epsilon m^2}{3}r\tau^2 \qdot
\end{equation}
We can substitute this expression in \myeqref{eq:flatKGtau} to get
\begin{equation}\label{eq:flatKGtauonly}
\tau''+\frac{2\tau'}{r}-\frac{(\tau')^2}{\tau}+\frac{\epsilon^2 m^4}{9}r^2\tau^3+\tauexp m^2\tau=0 \qdot
\end{equation}
In terms of the dilaton $\varphi=\ln{\tau}$, the Klein-Gordon equation \myeqref{eq:KGphi} then becomes
\begin{equation}\label{eq:flatKGphi}
\varphi''+\frac{2}{r}\varphi'+\frac{\epsilon^2 m^4}{9}r^2 e^{2\varphi}+\tauexp m^2=0 
\end{equation}
where we impose the boundary equation $\varphi'(0)=0$.

\subsubsection{Perturbative expansion in \texorpdfstring{$\epsilon$}{TEXT}}
Up to now, everything is exact. We now expand in powers of $\epsilon$:
\begin{align}
\varphi & =\varphi^{(0)}+\epsilon\varphi^{(1)}+\dots \qcom\\
a &= a^{(0)}+\epsilon a^{(1)}+\dots \qcom
\end{align}
and impose the boundary conditions at each order.

The advantage of this perturbative method is that the problematic term $e^{2\varphi}$ in \myeqref{eq:flatKGphi} appears only at second order in $\epsilon$. Indeed we get at orders 0 and 1
\begin{align}
{\varphi^{(0)}}''+\frac{2}{r}{\varphi^{(0)}}' +\tauexp m^2 & = 0 \qcom\\
{\varphi^{(1)}}''+\frac{2}{r}{\varphi^{(1)}}' & = 0 \qdot
\end{align}
The solution for $\varphi^{(1)}$ is then
\begin{equation}
    \varphi^{(1)}=c_0+\frac{c_1}{r} \qdot
\end{equation}
For $\varphi^{(0)}$ we have
\begin{equation}
    \varphi^{(0)}=d_0+\frac{d_1}{r}-\frac{\tauexp m^2}{6}r^2 \qdot
\end{equation}
The boundary conditions ${\varphi^{(n)}}'(0)=0$, impose that the fields are regular at 0, and so we have no $1/r$ term, i.e. $c_1=d_1=0$, and finally
\begin{equation}
\tau=(\tau_0+\epsilon\tau_1)e^{-\frac{\tauexp m^2}{6}r^2}+\dots
\end{equation}
where we have redefined the constants of integration.
Inside the source and for small $\epsilon$, $\tau$ decreases exponentially fast.

We can now  obtain $a$. Using \myeqref{eq:afromtau} we have
\begin{equation}
  {a^{(0)}}'+\epsilon {a^{(1)}}'= \epsilon\frac{ \tau_0^2}{2}\left(e^{-\frac{\tauexp m^2}{3}r^2}\right)' \qcom 
\end{equation}
and therefore
\begin{align}
a^{(0)} &= \text{cste} \qcom\\
a^{(1)} & = \frac{\tau_0^2}{2}e^{-\frac{\tauexp m^2}{3}r^2}+\text{cste} \qdot
\end{align}
The axion field is then given by 
\begin{equation}
a=a_0+\epsilon\left(a_1+\frac{\tau_0^2}{2}e^{-\frac{\tauexp m^2}{3}r^2}\right)+\dots \qdot
\end{equation}
We see that the axion and dilaton  fields only evolve when $\epsilon\ne 0$.

\subsubsection{Matching to the exterior solution}
The continuity of $\tau$ and $a$ at $r=R$ reads
\begin{align}
(\tau_0+\epsilon\tau_1)e^{-\frac{\tauexp m^2}{6}R^2} &= \frac{\beta}{\cosh(\frac{\beta\gamma}{R}+\delta)} \qcom\label{eq:contRtau} \\
a_0+\epsilon\left(a_1+\frac{\tau_0^2}{2}e^{-\frac{\tauexp m^2}{3}R^2}\right) &= \alpha-\beta\tanh(\frac{\beta\gamma}{R}+\delta) \qdot\label{eq:contRa}
\end{align}
We use the continuity equations for $\varphi'=\tau'/\tau$ and $a'/\tau^2$ 
\begin{align}
-\frac{\tauexp m^2}{3}R &= \tanh(\frac{\beta\gamma}{R}+\delta)\frac{\beta\gamma}{R^2} \qcom\label{eq:contRdphi} \\
-\frac{\epsilon\tauexp \tau_0^2 m^2}{3(\tau_0+\epsilon\tau_1)^2}R &= \frac{\gamma}{R^2} \qdot\label{eq:contRda_tau2}
\end{align}
We have thus eight integration constants, i.e. $\alpha$, $\beta$, $\gamma$, $\delta$ for the exterior solution and $\tau_0$, $\tau_1$, $a_0$, $a_1$ for the interior solution. Recall that
\begin{align}
\gamma & =-\frac{\epsilon m^2}{3}R^3 \qcom\\
\gamma\alpha & = -m^2 \int_0^R dr r^2 (\tauexp+\epsilon a(r))\qdot
\end{align}
So $\gamma$ is fixed independently of the others and $\alpha$ depends on $a_0$ and $a_1$. We are left with 6 parameters. With a system of 4 continuity constraints as given above, we end up with 2 degrees of freedom. These can be parameterised by the values of the fields at infinity $\tau_\infty$ and $a_\infty$ which determine the full solution.

\subsection{Screening revisited}\label{sec:reevaluating}
We can now revisit the conditions under which  the gravitational deviation from GR in the Jordan frame is small. Using 
\begin{equation}
\gamma\alpha=-\frac{1}{3M_p^2}\int_0^R dr r^2 (\tauexp\rho(r)+a(r)\cA(r))\qcom
\end{equation}
and  $\gamma=-{\epsilon m^2R^3}/{3}$ we obtain the identity
\begin{eqnarray}
\frac{-\epsilon m^2R^3}{3}\alpha=-m^2\left(\tauexp\frac{R^3}{3} +\epsilon\int_0^R dr r^2a^{(0)}(r)\right. \nonumber\\ +\left.\epsilon^2\int_0^R dr r^2a^{(1)}(r)\right) \qdot
\end{eqnarray}
As a result, the expansion of $\alpha$ is singular in the limit $\epsilon \ll 1$ and becomes
\begin{equation}
\alpha=\frac{\tauexp}{\epsilon}+a_0+\epsilon a_1 +\dots 
\end{equation}
implying that the outside solution is very sensitive to small values of $\epsilon$. In particular, we see that the limit of small $\epsilon \ll 1$ leads to a large value for the exterior axion field. Using \myeqref{eq:BQ_infty} we obtain 
\begin{equation}
\beta=\sqrt{\tau_\infty^2+\left(\frac{\tauexp}{\epsilon}+a_0-a^{(0)}_\infty\right)^2}
\end{equation}
where we have neglected the terms of order $\epsilon$. Now unless $a_\infty$ turns out to be of order $1/\epsilon$ and cancels exactly the term in $\kappa/\epsilon$, we find that for generic boundary values at infinity
\begin{equation}\beta=\frac{\tauexp}{\epsilon}+a_0-a_\infty^{(0)}+\dots\end{equation}
when $\kappa={\cal O}(1)$ and $\epsilon \ll 1$.
The matching conditions at $r=R$ simplify as we notice that $m^2 R^2/2 = G_N M/R $ where $M$ is the mass of the object. This is nothing but the Newtonian potential of the compact object at its surface which is always small in our Newtonian approximation, its value being close to $10^{-6}$ for the Sun. 
We then deduce using (\ref{eq:contRa}) that 
 $a_0=a_\infty^{(0)}$ to leading order and similarly
\begin{equation}
\tanh{\delta}=1- \frac{\epsilon^2}{\kappa}\frac{\tau_0^2}{2}+\dots
\end{equation}
implying that $\delta $ is always large. 
As a result we have
$
\epsilon\beta\tanh{\delta}=\tauexp+\dots \qdot$
and the PPN parameters are simply the ones of a scalar tensor theory with a coupling $\kappa/\sqrt 6$
\begin{equation}\gamma_{PPN}=\frac{3-\tauexp^2}{3+\tauexp^2}+{\cal O}(\epsilon)\qcom\end{equation}
and 
\begin{equation}\beta_{PPN}=1+{\cal O}(\epsilon^2)\qdot\end{equation}
In this limit  $\beta_{PPN}$ can be  arbitrarily close to $1$ for small $\epsilon$, but not $\gamma_{PPN}$. Small deviations of $\gamma_{PPN}$ from unity are only achieved for very small $\kappa$, a result which does not differ from  Brans-Dicke's and signals that screening does not take place. Screening can only take place when
\be
a_\infty^{(0)}-a_0=\frac{\kappa}{\epsilon}.
\ee
which corresponds to a specific choice for the axion field at infinity. As the theory has no scalar potential, the value of the axion field at infinity is not obtained by minimising an effective potential like in the chameleon mechanism. Hence, the boundary value of the axion field must emerge from the cosmological dynamics. We will study this below.

Our analysis has assumed that  $\cA=\epsilon\rho$. As soon as the dependence of the matter action on the axion is weak, i.e. $\cA/\rho\sim\epsilon\ll 1$,  the same qualitative results follow  as $\gamma$ will be of order $\epsilon$ and  $\gamma\alpha$ of order $\kappa$.
This reasoning is independent of the details  of the model inside the source. 
\subsection{The effective metric}

As we have seen, the generic absence of screening leads to the coupling of a compact of object to be equivalent to the one of a point particle with coupling $\kappa/\sqrt 6$. This is the coupling of matter to the dilaton.  The fact that the axion couples to the matter action too implies that compact objects do not follow the geodesics of the Jordan metric but the ones of an effective metric whose presence can be inferred from the small field expansion
\be 
\delta S_m= -\int d^4x \sqrt{-g}\left(\frac{\partial \ln A}{\partial \varphi}\delta \varphi  - \frac{\epsilon}{2}\delta a\right) T
\ee
where the variation of the fields is taken around the background values for the dilaton and the axion.
Notice that the axion and the dilaton fields both couple to the trace of the energy momentum tensor. Let us define 
\be 
B(\varphi,a)= A(\varphi) e^{-\epsilon a /2}
\ee
then the coupling to matter can be written as 
\be 
\delta S_m= -\int d^4x \sqrt{-g}\left(\frac{\partial \ln B}{\partial \varphi}\delta \varphi  +\frac{\partial \ln B}{\partial a}\delta a\right) T
\ee
corresponding to the coupling of a two-field scalar-tensor theory where the effective metric is
\be 
g_{\mu\nu}^{\rm eff}= B^2(\varphi,a) g_{\mu\nu}.
\ee
As a result, compact objects evolve along the geodesics of the effective metric and not the Jordan metric. 

This can be confirmed by analysing the geodesic equations for pressureless matter for the axio-dilaton theories. Indeed the Klein-Gordon equations and  the Bianchi identity $\nabla^\mu (R_{\mu\nu}-\frac{R}{2} g_{\mu\nu})=0$ imply the non-conservation equation
\begin{equation}
    \nabla_\mu {T}^{\mu\nu}=\frac{1}{2}(\tauexp T\partial^\nu\varphi-\cA\partial^\nu a)\qdot
\end{equation}
For non-relativistic matter, the energy momentum tensor is simply
\begin{equation}
{T}_{\mu\nu}=\rho u_\mu u_\nu \qwith u_\mu u^\mu=-1 \qdot
\end{equation}
In this section, dots will be denoting the time derivative along the particle lines  defined by $u^\mu$, i.e. 
$
    \dot{X}\equiv \nabla_u X=u^\mu\nabla_\mu X \qdot
$
For scalar quantities $X$, this also corresponds to the derivative with respect to the proper time of a particle  moving with velocity $u^\mu$. We define the local Hubble rate as
$
    3h\equiv \nabla_\mu u^\mu \qdot
$
The non-conservation equation then becomes
\begin{equation}\label{eq:CPT_NL_matter}
    \dot{\rho}u^\mu+3h\rho u^\mu=\frac{1}{2}(\tauexp\rho\partial^\mu\varphi+\cA\partial^\mu a)\qdot
\end{equation}
Contracting with $u_\mu$ and using $u^\mu u_\mu =-1$ we get the generalised continuity equation
\begin{equation}\label{eq:CPT_general_continuity}
    \dot{\rho}+3h\rho=-\frac{1}{2}(\tauexp\rho\dot{\varphi}+\cA\dot{a})\qdot
\end{equation}
We recognise the coupling function $B$ 
\begin{equation}
    B \equiv e^{-\frac{1}{2}(\tauexp{\varphi}+\epsilon{a})}
\end{equation}
when $\epsilon=\text{cste}$. We can define a conserved density $\rho_\text{con}$ in the Einstein frame such that 
\be 
\rho=B\rho_\text{con}
\ee
as
\begin{equation}\label{eq:CPT_con_density}
    \dot{\rho}_\text{con}+3h\rho_\text{con}=0\qdot
\end{equation}
This is the conserved matter density in the axion-dilaton setting. 
Combining \myeqref{eq:CPT_NL_matter} and \myeqref{eq:CPT_general_continuity}, we obtain the modified Newton's Law
\begin{equation}
    \dot{u}^\mu-\frac{1}{2}(\tauexp\dot{\varphi}+\epsilon\dot{a})u^\mu=\frac{1}{2}(\tauexp\partial^\mu{\varphi}+\epsilon\partial^\mu{a})\qdot
\end{equation}
This reads 
\be 
\dot{u}^\mu + \frac{d\ln B}{d\eta} u^\mu = -\partial^\mu \ln B.
\ee
where $\eta$ is the proper time. Defining by $m_0$ the mass of the particles, we find that the effective mass of these particles in the Einstein frame is dressed by the scalar field and becomes
\be 
m=B(\varphi,a) m_0.
\ee
This follows from the identification $\rho= m \delta^{(4)}( x^\mu -x^\mu(\tau))$ along the particle's trajectory and $\rho_{\rm con}= m_0 \delta^{(4)}( x^\mu -x^\mu(\tau))$. 
The momentum of each particle becomes 
\be 
p^\mu= m u^\mu. 
\ee
Newton's law then becomes
\be
\dot p^\mu= - m\partial^ \mu \ln B.
\ee
As a result, a force deriving from the potential $\ln B$ is exerted on each particle whose mass is also field dependent. For instance in the non-relativistic limit and in the presence of gravity, Newton's law becomes
\be 
\frac{dp^i}{dt}= -m \partial^i \Phi
\ee
where $\Phi_N$ is Newton's potential and 
\be 
\Phi= \Phi_N + \ln B
\ee 
combines the effects of gravity and the scalar field. This modification of Newton's law is nothing but the one which can be derived from the coupling of matter to the effective metric $g_{\mu\nu}^{\rm eff}$. As a result we have confirmed that compact objects do not follow the geodesics of the Jordan metric but the one of the effective metric. We will analyse the cosmological consequences of this result below. 

\subsection{The effective charge}

Let us come back to the effective scalar charge carried by a compact object. We have seen that in the $\epsilon\ll 1$ limit and unless the fields at infinity take special values, which should be adjusted cosmologically, the objects are not screened. Far away from a given object we expect the acceleration of another object due to the scalar field to fall off as
\be 
a^i\simeq -\frac{2 Q^2 G_N M}{r^3}r^i
\ee
where $Q$ is the scalar charge of both objects. Here $M$ is the mass of the object responsible for the acceleration of the second body.   The charges of both objects  are equal as no screening takes place. 
Using 
\be 
\ln B= -\frac{1}{2}(\kappa \varphi + \epsilon a) \supset \frac{1}{2}(\kappa \ln \cosh X(r) + \epsilon \beta \tanh X(r))
\ee 
and identifying this to $-2Q^2 G_N M/r$ at large distance we find  that far away from the object 
\be 
2 Q^2 G_N M =- \frac{\gamma \beta}{2} (\kappa \tanh \delta + \epsilon \beta (1-\tanh^2 \delta))
\ee
and for $\epsilon \ll 1 $ we retrieve that 
\be 
Q= \frac{\kappa}{\sqrt 6}
\ee
up to corrections of order $\epsilon^2$. The resulting interaction including gravity is equivalent to rescaling Newton's constant as 
\be 
G_{\rm eff}=(1+2Q^2) G_N
\ee
with $\Phi= -G_{\rm eff} M/r$.
As expected in this limit corresponding to $\epsilon \ll \kappa$, the coupling of the axion field to matter becomes negligible for far-away objects and the coupling is the same as in the Jordan frame. 

In conclusion, we find that $\kappa \lesssim  10^{-3}$ for the Cassini test to be evaded. This is a very small value which could only be avoided if the cosmological values of the axion and dilaton fields were tuned cosmologically.

\subsection{Numerical integration}

In this section, we will find numerical solutions of the equations around a massive sphere. 
\subsubsection{Setting the numerical problem}
The Klein-Gordon equations with the constant source inside a ball of radius $R$ have  the form
\begin{equation}\tau''+\frac{2\tau'}{r}-\frac{(\tau')^2}{\tau}+\frac{(a')^2}{\tau}+\frac{\tauexp \rho_0 \tau}{3 M_p^2}\theta(r)=0 \qcom\end{equation}
\begin{equation}a''+\frac{2 a'}{r}-\frac{2 a' \tau'}{\tau}+\frac{\epsilon \rho_0 \tau^2 }{3 M_p^2}\theta(r)=0 \qcom\end{equation}
where $\theta(r)$ is the step function that goes from 1 to 0 at the radius of the source $R$. We introduce the characteristic length $L=\sqrt{3 M_p^2/\rho_0}=m^{-1}$, and write $r=\hat{r}L$.  We obtain the dimensionless equations:
\begin{equation}{\tau}''+\frac{2{\tau}'}{\hat{r}}-\frac{({\tau}')^2}{{\tau}}+\frac{({a}')^2}{{\tau}}+ \tauexp{\theta}(\hat{r}){\tau}=0 \qcom\end{equation}
\begin{equation}{a}''+\frac{2 {a}'}{\hat{r}}-\frac{2 {a}' {\tau}'}{{\tau}}+\epsilon {\theta}(\hat{r}){\tau}^2=0 \qcom\end{equation}
which we solve with the initial conditions ${\tau}'(0) = {a}'(0) = 0$.  We also regularise  the step function, e.g.: $\hat{\theta}(\hat{r})=\frac{1}{2}(\tanh(N\frac{\hat{R}-\hat{r}}{\hat{R}})+1)$ to have a transition of width $\sim \hat{R}/N$.
 Notice that $\hat{R}=R/L=\sqrt{2GM/R}$, i.e the square root of the ratio of the Schwarzschild radius to the radius of the source. For the Sun $\Rsun=2.05\times 10^{-3}$.

\subsubsection{Results}

\begin{figure}[ht!]
\centering
\includegraphics[width=0.5\textwidth]{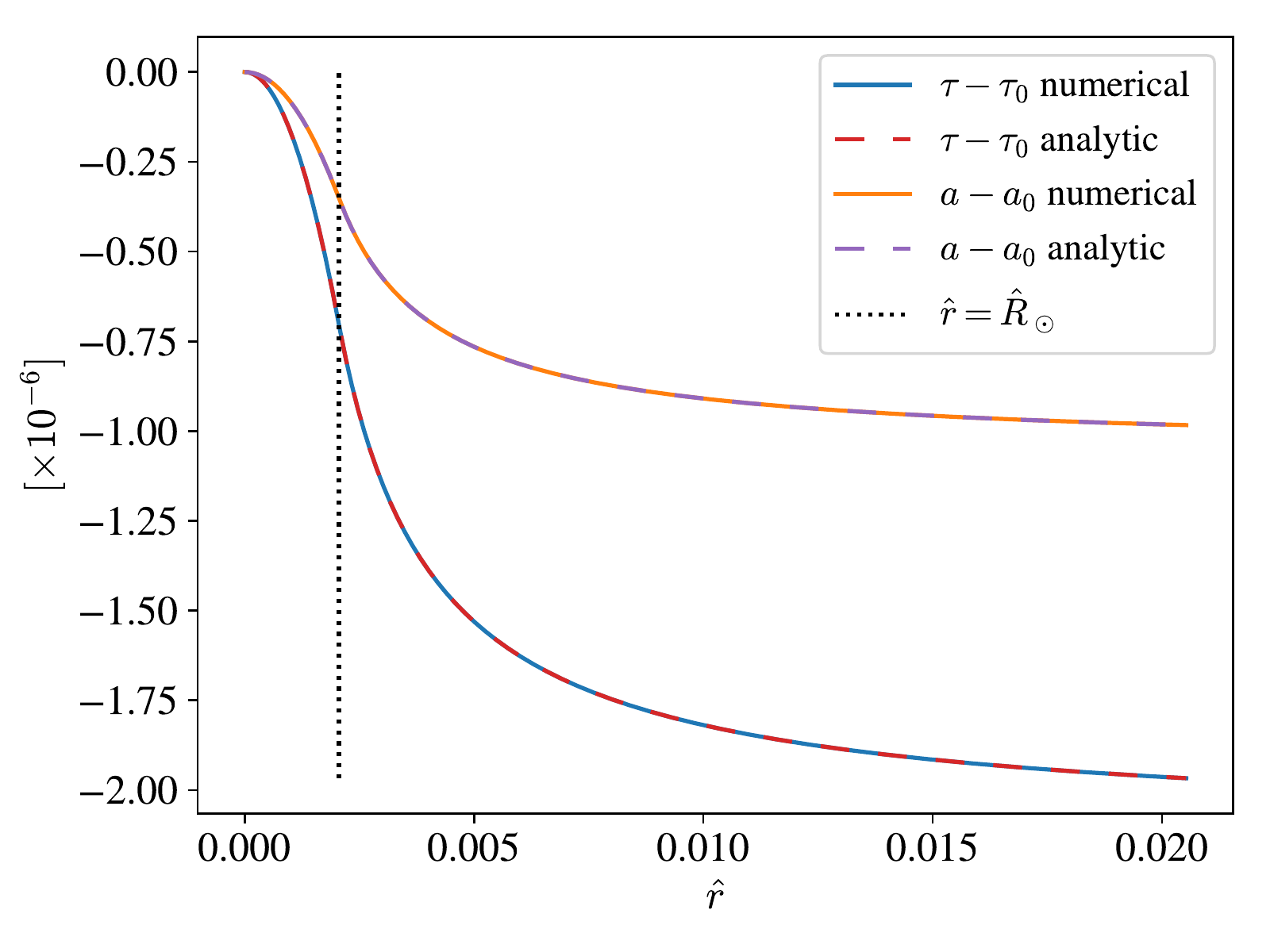}
\caption{The field profiles for $\hat{R}=\Rsun$, $\tauexp=1$ and $\epsilon=0.5$. The initial values are taken such that $\tau_0=a_0=1$, $\tau'(0)=a'(0)=0$.
}
\label{fig:fields_sun}
\end{figure}

We see in the example in \myfigref{fig:fields_sun} that we obtain a perfect match with the exterior solution, but also with the first order interior solution. The variations of the fields  are relatively small, i.e. of order $10^{-6}$ relative to the  value at the centre. For other sets of parameters the variation increases with the initial value at the centre and of course the value of $\kappa$ and $\epsilon$. 

The equations that we have used in the flat metric approximation are valid only for
\begin{equation}
r\gg G M \quad , \quad r\gg \vert\gamma\beta \vert \qdot
\end{equation}
With $R=\hat{R}L$, we have
\begin{equation}
r\gg GM \iff \hat{r}\gg\frac{\hat{R}^3}{2} \qdot
\end{equation}
For the Sun, 
$\Rsun^3\sim 10^{-8}$.
One can trust our description of the static solution apart from a small region around the origin.
Similarly, we have $\gamma= -\epsilon \hat{R}^3 L / 3$, so we get
\begin{equation}r\gg\vert\gamma\beta\vert \iff \hat{r} \gg  \hat{R}^3 \frac{\epsilon \beta}{3} \qcom\end{equation}
which excludes a very small region around the origin.

\section{Axio-dilaton cosmology}\label{sec:bcosmology}

\subsection{Two fluid model}\label{sec:two_fluids}

In the following, we will concentrate on models where two fluids are present, i.e. the baryons and Cold Dark Matter (CDM) with couplings determined by $\kappa$ to the dilaton and $\epsilon_{B,C}$ to the axion. We have seen that $\kappa$ and $\epsilon_B$, i.e. the coupling of the baryons to the axion, must be small to comply with the solar system tests unless the fields take special values at infinity. This translates into a choice of boundary conditions for the fields on cosmological scales now. 
We will see that this situation is not generic and that starting from initial conditions in the radiation era which do not perturb the background cosmology, the cosmological dynamics do not drive the fields to special values now.

The energy-momentum  tensor of matter is  taken to be   
\begin{equation}
\munu{T}^{B,C}=\rho_{B,C} u_\mu^{B,C} u_\nu^{B,C}
\end{equation}
corresponding to the baryons $B$ and CDM  $C$. 
We will also assume here for definiteness
\begin{equation}
\cA=-\epsilon_C T_C - \epsilon_B T_B
\end{equation}
where $T_{B,C}=-\rho_{B,C}$ are the baryonic and  CDM densities respectively. 
The matter energy momentum tensors are not conserved but satisfy 
 the non-conservation equations
\begin{equation}
    \nabla_\mu {T}^{\mu\nu}_i=\frac{1}{2}(\tauexp T_i\partial^\nu\varphi+ \epsilon_i T_i\partial^\nu a)
\end{equation}
where $i=B,C$. This implies that the total energy momentum tensor
\be 
T^{\mu\nu}=T_B^{\mu\nu}+ T^{\mu\nu}_C
\ee
satisfies the non-conservation equation
\begin{equation}
    \nabla_\mu {T}^{\mu\nu}=\frac{1}{2}(\tauexp T\partial^\nu\varphi-\cA\partial^\nu a)
\end{equation}
which is a consequence of the Bianchi identity.

In this section, we will denote the time derivative along the particle lines,  defined by $u^\mu_i$, by $d/d\tau_i= u^\mu_i \nabla_\mu$
We define the local Hubble rate as
$
    3h_i\equiv \nabla_\mu u^\mu_i \qdot
$
Notice that the covariant derivatives are calculated in the Einstein frame, hence this is the local Hubble rate along the particle lines as measured using the geometry of the Einstein frame. 
The non-conservation equations for each species then become
\begin{equation}\label{eq:CPT_general_continuity_2fluids}
    \frac{d\rho_i}{d\tau_i}+3h_i\rho_i=-\frac{1}{2}(\tauexp\dot{\varphi}+\epsilon_i\dot{a})\rho_i\qdot
\end{equation}
We define  the coupling function $B_i$ 
\begin{equation}
    B_i \equiv e^{-\frac{1}{2}(\tauexp{\varphi}+\epsilon_i{a})}
\end{equation}
when $\epsilon_i=\text{cste}$. We can now introduce  a conserved density $\rho_\text{con,i}$ in the Einstein frame such that 
\be 
\rho_i=B_i\rho_\text{con,i}
\ee 
and 
\begin{equation}\label{eq:CPT_con_density_2fluids}
    \frac{d\rho_{{\rm con},i}}{d\tau_i}+3h_i\rho_{\rm{con},i}=0\qdot
\end{equation}
This is the conserved matter density in the axio-dilaton setting. 
Similarly we obtain the modified Newton's Law
\begin{equation}
    \frac{d{u}^\mu_i}{d\tau_i}+ \frac{d\chi_i}{d\tau_i}u^\mu_i=-\partial^\mu \chi_i
\end{equation}
where
\be 
\chi_i\equiv \ln B_i.
\ee
For each species, we can define an effective metric
\be 
g^i_{\mu\nu}= B_i^2 g_{\mu\nu}
\ee
which corresponds to the Jordan frame for the given species. As $B_B\ne B_C$, we see that the Jordan frames for CDM and the baryons do not coincide. 

We will apply this formalism first to the background cosmological case and then to the cosmological perturbations. 

\subsection{Spatially flat cosmology}

We are interested in  the cosmology of a homogeneous and isotropic Universe in the presence of the  axio-dilaton fields. The FLRW (Friedmann-Lema\^itre-Robertson-Walker) metric reads
\begin{equation}
\begin{aligned}
{g}_{\mu\nu}d x^\mu d x^\nu &= -d t^2+R^2(t)\gamma_{ij}dx^i dx^j \\
&= - dt^2+R^2(t)(\frac{ dr^2}{1-k r^2}+r^2  d\Omega^2)
\end{aligned}
\end{equation}
where  $R$ is the scale factor. We define as usual the Hubble rate as $H=\dot{R}/R$. In the following we will focus on the spatially flat case $k=0$. 

We assume that  the fields are irrelevant in the early  Universe up until some redshift $z_i$ which will be typically the matter-radiation equality. Indeed, in the radiation era the matter density is negligible and therefore the fields are hardly influenced by their matter couplings. As a result they remain constant if their initial velocities vanish. This also guarantees that the influence of the axio-dilaton system on Big Bang Nucleosynthesis (BBN) is minimal. We will choose  the initial conditions for  the fields $\varphi$ , $a$ and their derivatives such that they vanish at $z_i$. 
This does not correspond to the values of the fields for which screening takes place. On the other hand, this entails a vanishing energy density for the axion and dilatons initially. 

Each species has its own Jordan frame with the background metric
\be
g_{\mu\nu}^i dx^\mu dx^\nu= -dt_i^2 + R^2_i dx^i dx_i
\ee
where the cosmic time and the scale factor are defined by
\be 
dt_i = B_i dt, \ \ R_i= B_i R.
\ee
As CDM is not subject to the precision tests in the solar system, 
we will allow for large values of $\epsilon_C$.
Moreover we will consider the effects on  the scale factor in the Jordan  frame of the baryons $R_B=B_B R$ corresponding  to the conservation of baryonic matter. We will use the convention that $R_B = 1$ today and identify the redshift as detected from transition lines of atoms by
\be 
1+z=R_B^{-1}.
\ee
Although we will focus on the dynamics in the baryon frame, we will first study the equations of motion in the Einstein frame. 

\subsection{The Klein-Gordon equations}
We look for time-dependent solutions for the scalar fields $\tau(t)$ and $a(t)$ of the Klein-Gordon equations
\begin{equation}\label{eq:cosmoKGtau}
\ddot{\tau}+3H\dot{\tau}-\frac{\dot{\tau}^2-\dot{a}^2}{\tau}-\frac{\tauexp\tau\rho}{3M_p^2}=0 \qcom
\end{equation}
and 
\begin{equation}\label{eq:cosmoKGa}
\ddot{a}+3H\dot{a}-\frac{2\dot{\tau}\dot{a}}{\tau}-\frac{\tau^2\cA}{3M_p^2}=0 \qdot
\end{equation}
Here the total matter density is  $\rho= \rho_B +\rho_C$ and ${\cal A}= \epsilon_B \rho_B + \epsilon_C \rho_C$. 
Notice that these equations are valid both in the radiation and matter eras. In the radiation era, the source terms depend on the matter density only as the trace of the radiation energy momentum tensor vanishes. As a first approximation, we will neglect the source terms in the radiation era. This implies that $\dot a \approx 0$ and $\dot \phi \approx 0$ and the field hardly move during the radiation era. In our numerical analysis, we will be interested in the physics in the matter era and will fix the initial conditions at matter-radiation equality. A detailed analysis of the model from the end of inflation through the radiation to the matter era is left for future work. We will consider that the fields start evolving significantly when the matter era begins.

\subsection{The continuity equations}
The energy density and pressure carried by the two scalar fields are given by
\begin{equation} \rho_f= \frac{3M_p^2}{4}(\frac{\dot{\tau}^2+\dot{a}^2}{\tau^2}) \qcom\end{equation}
and 
\begin{equation}
p_f= \frac{3M_p^2}{4}(\frac{\dot{\tau}^2+\dot{a}^2}{\tau^2}) \qcom
\end{equation}
corresponding to a  perfect fluid with equation of state $\omega_f=1$.
Using the Bianchi identity and the Einstein equation in the Einstein frame, we obtain that the total energy is conserved, i.e. 
\begin{equation}
\dot{\rho}+\dot{\rho}_f+3H(\rho+\rho_f+P+P_f)=0 \qcom
\end{equation}
where the pressure terms have been included. This leads to 
\begin{equation}\label{eq:continuity}
\dot{\rho}+\dot{\rho}_f+3H(\rho+2\rho_f)=0 \qdot
\end{equation}
 Using the identity
\begin{equation}
\dot{\rho}_f=\frac{3M_p^2}{4}\frac{2}{\tau^3}(\tau\dot{\tau}\ddot{\tau}+\tau\dot{a}\ddot{a}-\dot{\tau}^3-\dot{\tau}\dot{a}^2)
\end{equation}
and the  expression of $\ddot{\tau}$ and $\ddot{a}$ in  the Klein-Gordon equations \myeqref{eq:cosmoKGtau} and \myeqref{eq:cosmoKGa}, we get
\be
\dot{\rho}_f+ 6H\rho_f=\frac{1}{2}\left(\tauexp\rho\frac{\dot{\tau}}{\tau}+\cA\dot{a}\right).
\ee
Finally using the continuity equation we obtain
\begin{equation}\label{eq:continuity_simp}
\dot{\rho}+3H\rho+\frac{1}{2}(\tauexp\rho\dot\varphi+\cA\dot{a})=0
\end{equation}
as we have argued previously. This is associated to the non-conservation equations per species
\begin{equation}\label{eq:continuity_simpi}
\dot{\rho_i}+3H\rho_i+\frac{1}{2}\rho_i(\tauexp\dot\varphi+\epsilon_i\dot{a})=0.
\end{equation}
This can be integrated exactly and leads to
\begin{equation}\label{eq:cosmo_rho}
\rho = \rho_B+\rho_C= B_B \frac{\rho_{0B}}{R^3}+ B_C \frac{\rho_{0C}}{R^3}.
\end{equation}
Notice that if we define 
\be
\rho= B \frac{\rho_0}{R^3}
\label{eq:rho_sol}
\ee
then
$
B= \lambda_B B_B + \lambda_C B_C
$
where the fraction of baryons and CDM are
$
\lambda_i=\frac{\rho_{0i}}{\rho_0}
$
where $\rho_0= \rho_{0B} + \rho_{0C}$. In practice we have 
$\lambda_B= \frac{\Omega_{0B}}{\Omega_{0B}+\Omega_{0C}}$ where $\Omega_{0B}\simeq 0.022$ and $\Omega_{0C} \simeq 0.12$ from the Planck experiment \cite{Planck:2018vyg}. Here and in the following we normalise the density $\rho_0$ to the Planck value as deduced from early time physics compared to the late time effects on the matter density that we will study below. 

\subsection{The Friedmann equations}

The Friedmann equation is obtained from the $(00)$ component of the Einstein equation and becomes
\begin{equation}\label{eq:fried1}
H^2=\frac{\rho}{3M_p^2}+\frac{\rho_f}{3M_p^2}+\frac{\rho_\Lambda}{3M_p^2}\qcom
\end{equation}
where we have introduced a cosmological constant associated to the energy density $\rho_\Lambda$.
The Raychaudhuri  equation from the $(ii)$ Einstein equation reads
\begin{equation}\frac{\ddot{R}}{R}=-\frac{1}{6M_p^2}(\rho+\rho_f+\rho_\Lambda+3(P+P_f+P_\Lambda)) \qcom\end{equation}
leading to 
\begin{equation}\label{eq:fried2}
\frac{\ddot{R}}{R}=-\frac{1}{6M_p^2}(\rho+4\rho_f-2\rho_\Lambda)\qdot
\end{equation}
These equations are defined in the Einstein frame and will be transformed into the effective frame for baryons below.

We thus have the following system of differential equations for $R$, $\tau$ and $a$ only:
\begin{equation}\label{eq:cosmo_system}
\begin{aligned}
\ddot{\tau}+3H\dot{\tau}-\frac{\dot{\tau}^2-\dot{a}^2}{\tau}-{\tauexp}\frac{\rho(\tau,a)}{3M_p^2}\tau &=0 \qcom \\
\ddot{a}+3H\dot{a}-\frac{2\dot{\tau}\dot{a}}{\tau}-\epsilon \frac{\rho(\tau,a)}{3M_p^2}\tau^2 &=0 \qcom \\
\dot{R}=H(\tau,a) R \qcom
\end{aligned}
\end{equation}
with the Friedmann equation
\begin{eqnarray}
&&H^2=\frac{\rho}{3M_p^2}+\frac{\rho_f}{3M_p^2}+\frac{\rho_\Lambda}{3M_p^2}\qcom\nonumber\\ 
&&\rho=B_B \frac{\rho_{0B}}{{R^3}}+B_C \frac{\rho_{0C}}{{R^3}},\ \rho_f=\frac{3M_p^2}{4}\left(\frac{\dot{\tau}^2+\dot{a}^2}{\tau^2}\right)\qcom\nonumber \\
&& \frac{d\ln B_i}{dt}= -\frac{1}{2}(\kappa \dot\varphi + \epsilon_i \dot a)\qdot\nonumber \\
\end{eqnarray}
We will solve these equations numerically for different values of $\epsilon$ and $\kappa$.

\subsection{Dynamics in the effective baryon frame}

In the baryon frame, the  Hubble rate is 
\begin{equation}H_B\equiv \frac{ d\ln R_B}{dt_B}=\frac{H}{B_B}+ \frac{d \chi_B}{dt_B}\qdot\end{equation}
The conserved baryon density in the baryon frame is simply
\be 
\tilde \rho_{B}\equiv \rho_{{\rm con}B} =  \frac{\rho_{0B}}{R_B^3}.
\ee
In this frame, CDM is not conserved but an observer fitting the evolution of the Universe with a prior that CDM is also conserved in the same frame as the baryons would identify  the conserved CDM density as 
\be 
\tilde \rho_C=  \frac{\rho_{0C}}{R_B^3},
\ee
and would write an  effective Friedmann equation in the baryonic frame 
\begin{equation}
H_B^2\equiv \frac{8\pi G_B}{3} \tilde\rho_B + \frac{8\pi G_C}{3} \tilde \rho_C +\frac{8\pi G_N}{3}\tilde \rho_{\Lambda}\qcom
\end{equation}
where we have used $8\pi G_N= 1/M_{\rm p}^2$.
This allows one to identify the effective Newton constants $G_{B,C}$ and the dark energy component $\bar \rho_\Lambda$. None of these constants are constant in the baryon frame. 
In practice, the effective Newton constants are determined by 
\be 
G_{B}=  B_B^2 {G_N},\ G_C=B_B B_C G_N
\ee
i.e. the two Newtonian constants evolve differently.
The dark energy component is simply defined as the complement to the baryon and CDM contributions in the baryonic Friedmann equation. 

As long as the fields do not evolve rapidly, i.e. at the beginning of the matter era we have $H_B\approx H/B_B$. 
The dark energy component becomes
\be 
\tilde \rho_\Lambda = \frac{ \rho_f +\rho_\Lambda}{B_B^2}.
\ee
In the late Universe, this identification is not valid anymore and a numerical integration of the equations of motion is necessary.

The same Friedmann equation in the baryon frame can be written as 
\be 
\Omega_B^B + \Omega_C^B+\Omega_\Lambda^B=1
\ee
where the energy fractions $\Omega^B_i$, $i=B,C,\Lambda$ are identified in the baryon frame and are such that
\be 
\Omega_{B,C}^B= \frac{8\pi G_{B,C} \tilde \rho_{B,C}}{3H_B^2}, \ \Omega_\Lambda^B= \frac{8\pi G_{N} \tilde \rho_{\Lambda}}{3H_B^2}.
\ee
The deviations of the energy fractions from $\Lambda$CDM are represented in Fig. \ref{fig:cosmo_proportions}.

\subsection{Deviations from \texorpdfstring{$\LCDM$}{TEXT} and observational constraints}

We will first define the effective gravitational constant that an observation would measure.  For example, Big Bang Nucleosynthesis (BBN) puts constraints on the variation of such a $G_{\rm eff}$ \cite{variation_G,funda_csts}:
\begin{equation}
 \abs{\Delta G/G}\equiv \abs{\frac{G_\text{eff}^{\text{today}}-G_\text{eff}^{\text{BBN}}}{G_\text{eff}^{\text{BBN}}}} <0.4~. 
\end{equation}
This  assumes a Hubble evolution similar to that of $\LCDM$ in the matter-dominated era. In our model, since observations are made in the baryon frame, the corresponding $G_\text{eff}$ satisfies
\begin{equation}H_B^2=\frac{8\pi}{3}G_\text{eff}\bqty{\frac{c_\rho}{R_B^3}+\rho_{\Lambda,B}}~, \label{Geff}\end{equation}
for some $c_\rho$. In order to  have $G_\text{eff}(z_i)=G_N$ initially, we set $c_\rho=\rho_0$, as defined by \myeqref{eq:rho_sol}, i.e we normalise Newton's constant by using the Planck normalisation in the early Universe. Since the physics between  BBN and $z=z_i$ is the same as in the standard model, $G_\text{eff}=G_N$ till $z=z_i$. Later the  relative variation of the effective coupling to baryons  can be computed between $z=z_i$ and today: 
\begin{equation}\eval{\frac{\Delta G_B}{G_B}}_{{\rm BBN}\rightarrow{\rm today}}=\eval{\frac{\Delta G_B}{G_B}}_{z_i\rightarrow{\rm today}}\qdot\end{equation}

We can further constrain the possible deviations of the Hubble rate from the standard model by imposing that this should be less than the discrepancy appearing in the $H_0$ tension. 
Indeed, there are two diverging determinations  of the present time Hubble rate $H_0$ with a relative difference of order $10\%$ \cite{hubble_tension}. In the axio-dilaton theory, the fact that Newton's constant varies implies that the Hubble now differs from the corresponding Hubble rate in the standard model . We have normalised the Hubble rates to coincide at the beginning of the matter era. This motivates looking for parameters that satisfy
\begin{equation}
    \abs{\frac{\Delta H_B}{H_B}}_{\rm tension}\equiv\abs{\frac{H_B(z=0)-H_{\rm SM}(z=0)}{H_{\rm SM}(z=0)}} <0.1\qcom
\end{equation}
where $H_{\rm SM }$ is the Hubble rate in the standard model.
Another stringent constraint comes from BAO (Baryon Acoustic Oscillations) \cite{eBOSS:2020yzd} which specify that the deviations of $H_B(z)$ for $0.2 \lesssim z\lesssim 2.5$ should less than around 3 percent \cite{eBOSS:2020yzd}
\begin{eqnarray}
 &&\abs{\frac{\Delta H_B}{H_B}}_{\rm BAO}\equiv \nonumber \\ && \abs{\frac{H_B(z\in [0.2,2.5])-H_B^{\rm SM}(z\in [0.2,2.5])}{H_B^{\text{SM}}(z\in [0.2,2.5])}}  <0.03\qdot  \nonumber \\
\end{eqnarray}
This implies that the differences between $\Lambda$-CDM and the axio-dilaton models must appear late in the evolution of the Universe. We will see that the BAO constraint is the most stringent one amongst the ones we have selected. Of course a much more precise numerical  study is required to constrain the parameter space. This is left to future work. 

We also  consider the effective equation of state of dark energy.  In  GR  with matter in addition to a fluid $X$ with equation of state $w$,  the deceleration parameter
\begin{equation}
    q_0:=-\abs{\frac{\ddot{R}}{R H^2}}_{\text{today}}\qdot
\end{equation}
is  given by
\begin{equation}\label{eq:q0}
    q_0=\frac{1}{2}(\Omega_{m,0}+(1+3\omega)\Omega_{X,0})
\end{equation}
where $\Omega_{i,0}=\rho_i(z=0)/3M_p^2H_0^2$. Observations give thus an estimate for $w$ depending on $\Omega_{m,0}$. For $\Omega_{m,0}\sim 0.3$, which can be obtained independently, this leads to  \cite{forEqofState,Peter_Uzan}
\begin{equation}
    w\sim -1\pm 0.1\qdot
\end{equation}
Recent constraints and future prospects can be found in \cite{Linder:2023klx}. We now define the effective equation of state
\begin{equation}\label{eq:weff_def}
    w_\mathrm{eff}\equiv \frac{1}{3}{\frac{2q_{B,0}-\Omega_{m,0}}{\Omega_{\Lambda,0}}-1}\qcom
\end{equation}
where:
\begin{equation}
    q_{B,0}=-\eval{\frac{\partial_{t_B}^2{R_B}}{R_B H_B^2}}_{\text{today}}\qdot
\end{equation}
Taking $\Omega_{m,0}\simeq 0.3$, we must impose the constraint
\begin{equation}
    \vert{\Delta w}\vert\equiv \vert{w_\mathrm{eff}+1}\vert \lesssim 0.1\qdot
\end{equation}
We will use this bound in what follows as a guiding principle. We are not trying to give a precise fit to the data but an indication on the parameter space compatible with cosmology.

\subsection{Numerical integration}
\subsubsection{Dimensionless equations}
In the following, we will use $L=\sqrt{3 M_P^2/\rho_0}$ as the unit of time and length. Here $\rho_0$ is defined via \myeqref{eq:cosmo_rho}. We will  obtain quantities as functions of redshift starting at matter-radiation equality.  We define  the system of dynamical equations with  the number of efolds $N$ and $\varphi$ such that $R=e^N$ and $\tau=e^{\varphi}$. We get
\begin{eqnarray}
&&\ddot{\varphi} = -3 \hat{H}\dot{\varphi}-\dot{a}^2 e^{-2\varphi}+{\tauexp}\hat{\rho}\nonumber \\
&&\ddot{a} = -3 \hat{H}\dot{a}+2\dot{\varphi}\dot{a}+\epsilon \hat{\rho} e^{2\varphi}\nonumber \\
&&\dot{N}=\hat{H}\nonumber \  \end{eqnarray}
with
\begin{eqnarray}
&&\hat{H}^2=\hat{\rho}+\hat{\rho}_f+\hat{\rho}_\Lambda,\quad \hat{\rho}=Be^{-3N}, \nonumber \\  &&\hat{\rho}_f=\frac{1}{4}(\dot{\varphi}^2+\dot{a}^2e^{-2\varphi})\qdot\nonumber \\
\end{eqnarray}
Here $\rhoL$ corresponds to $\rho_\Lambda/\rho_0$ where $\rho_\Lambda$ is the density associated to the cosmological constant. In $\LCDM$, $\rho_0$ is also the value of the matter density today, so that $\rhoL=\Omega_{\Lambda,0}/\Omega_{m,0}\simeq 7/3$ in $\LCDM$.
It is convenient to work in conformal time  $\eta$, such that $R d\eta = dt$, implying that
\begin{eqnarray}
&& {\varphi}'' = -2 \hat{\mathcal{H}}{\varphi}'-{a'}^2 e^{-2\varphi}+\tauexp\tilde{\rho}\nonumber \\
&& {a}'' = -2 \hat{\mathcal{H}}{a}'+2{\varphi'}{a'}+\epsilon \tilde{\rho} e^{2\varphi}\nonumber\\
&& {N}'=\hat{\mathcal{H}}\nonumber \\ \end{eqnarray}
with
\begin{equation}\hat{\mathcal{H}}^2=\check{\rho}+\check{\rho}_f+\check{\rho}_\Lambda \qcom\end{equation}
where the derivatives are now with respect to ${\eta}$.
We have rescaled all the densities as $\check{\rho}=e^{2N}\hat{\rho}$ and similarly for the scalar and dark energy parts. 

\subsubsection{Results}

\begin{figure*}
    \centering
    \begin{subfigure}[b]{0.45\textwidth}
        \centering
        \includegraphics[width=\textwidth]{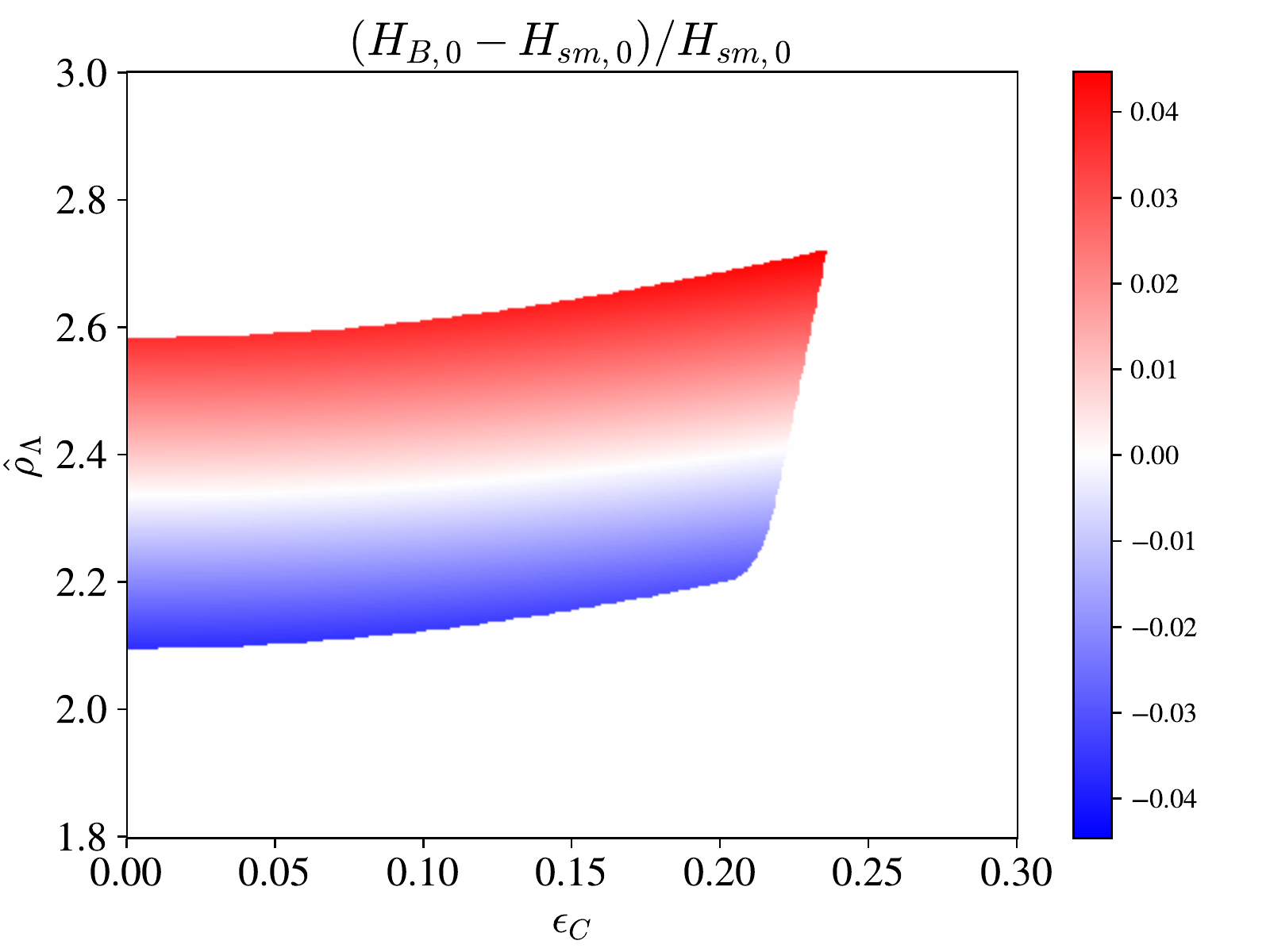}
        \label{fig:dHH_imshow_k0p001_deta0p0001}
    \end{subfigure}
    \begin{subfigure}[b]{0.45\textwidth}
        \centering
        \includegraphics[width=\textwidth]{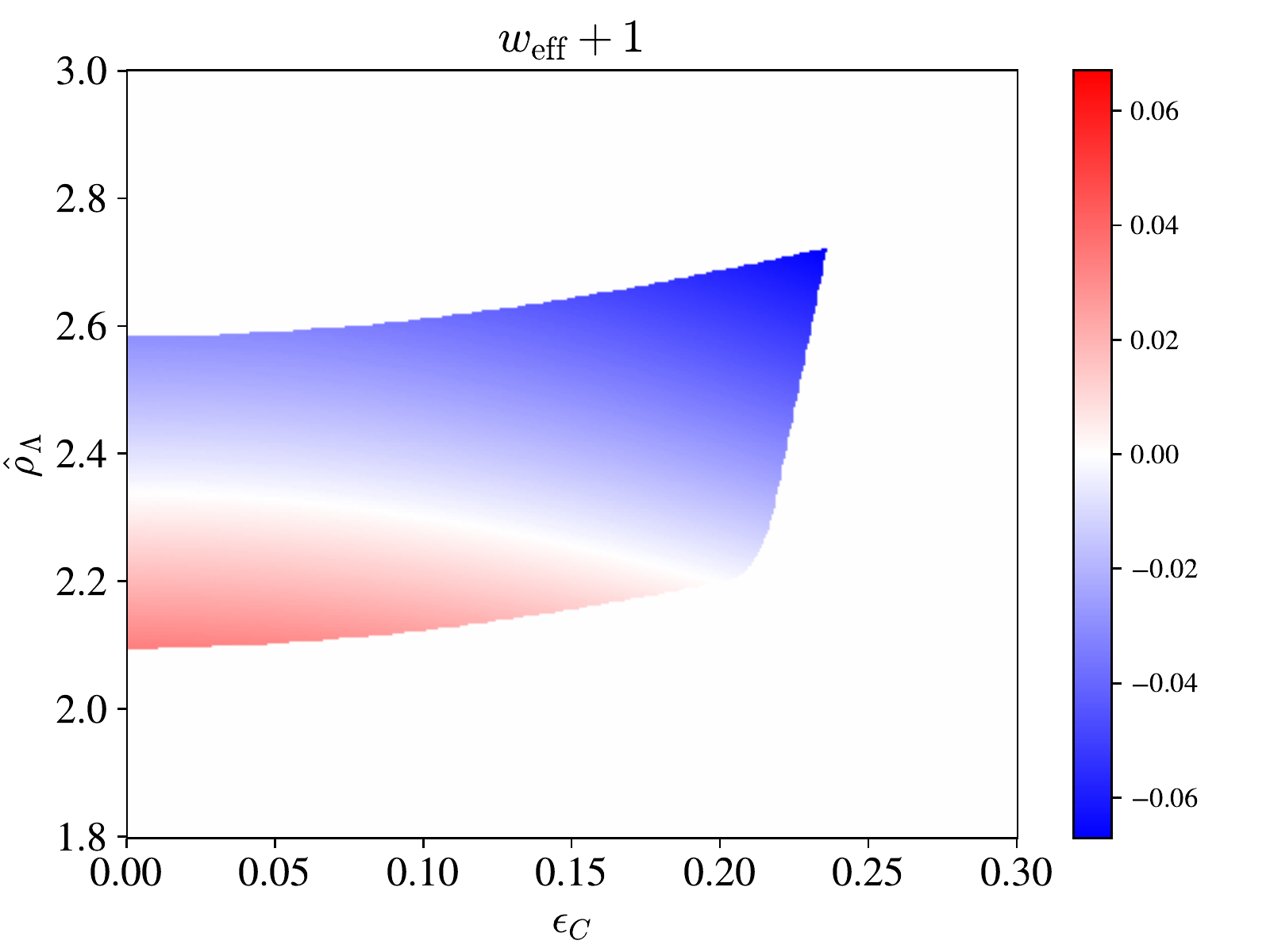}
        \label{fig:omega_eff_imshow_k0p001_deta0p0001}
    \end{subfigure}
    \begin{subfigure}[b]{0.45\textwidth}
        \centering
        \includegraphics[width=\textwidth]{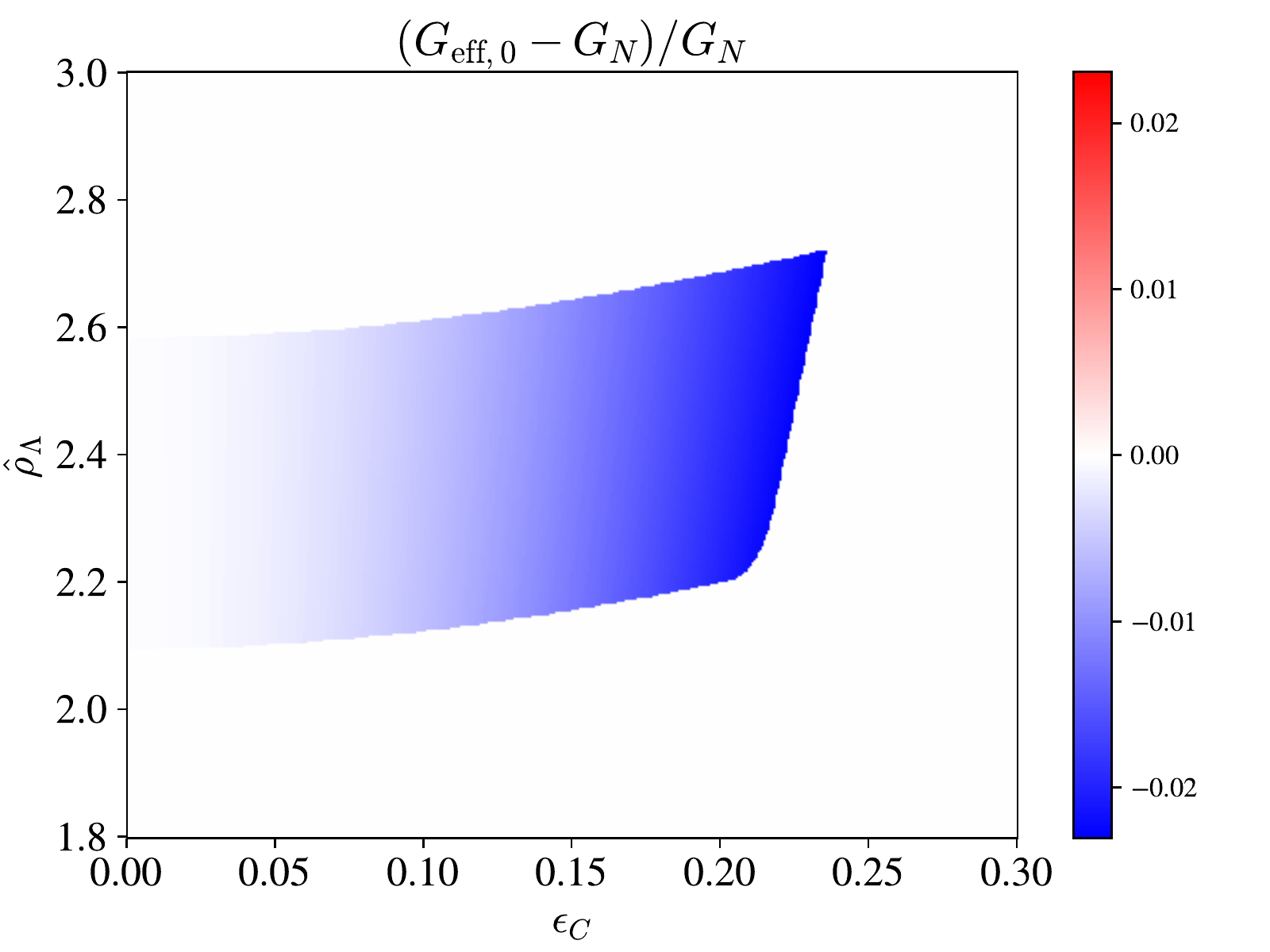}
        \label{fig:dGeff_imshow_k0p001_deta0p0001}
    \end{subfigure}
    \begin{subfigure}[b]{0.45\textwidth}
        \centering
        \includegraphics[width=\textwidth]{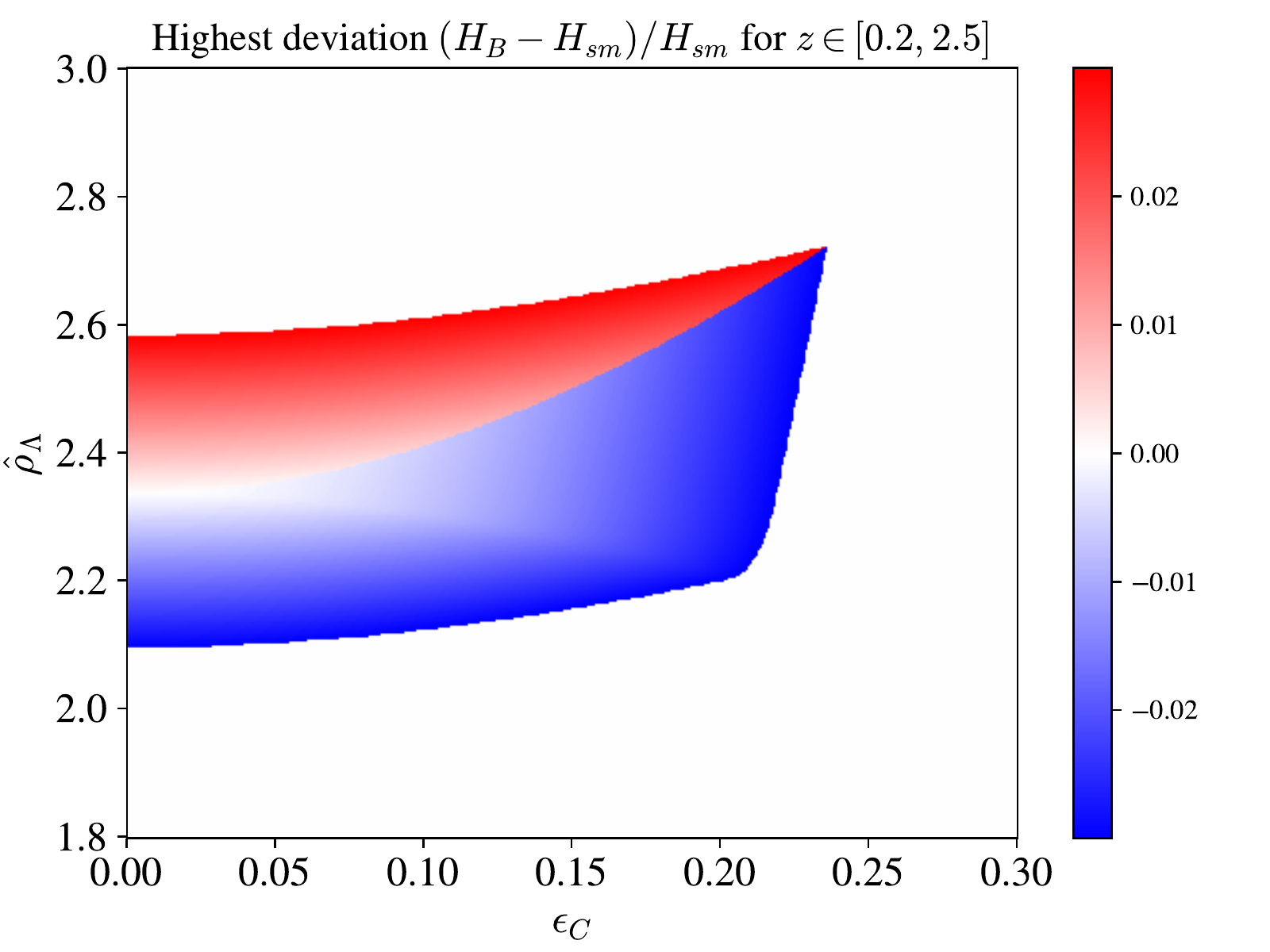}
        \label{fig:valid_vals_imshow_k0p001_deta0p0001}
    \end{subfigure}
    \caption{We show the values of the parameters which satisfy the cosmological tests  for $\tauexp=10^{-3}$ and $\epsilon_B=10^{-3}$. $602\times 602$ points are plotted. The coloured regions satisfy the four constraints: $\vert{\Delta H_0/H_0}\vert< 0.1$, $\vert{\Delta w}\vert<0.1$, $\vert{\Delta G/G}\vert<0.4$, $|\Delta H/H|_\text{BAO}<0.03$.}
    \label{fig:Observables_k0p001}
\end{figure*}
\begin{figure*}
    \centering
    \includegraphics[width=0.6\textwidth]{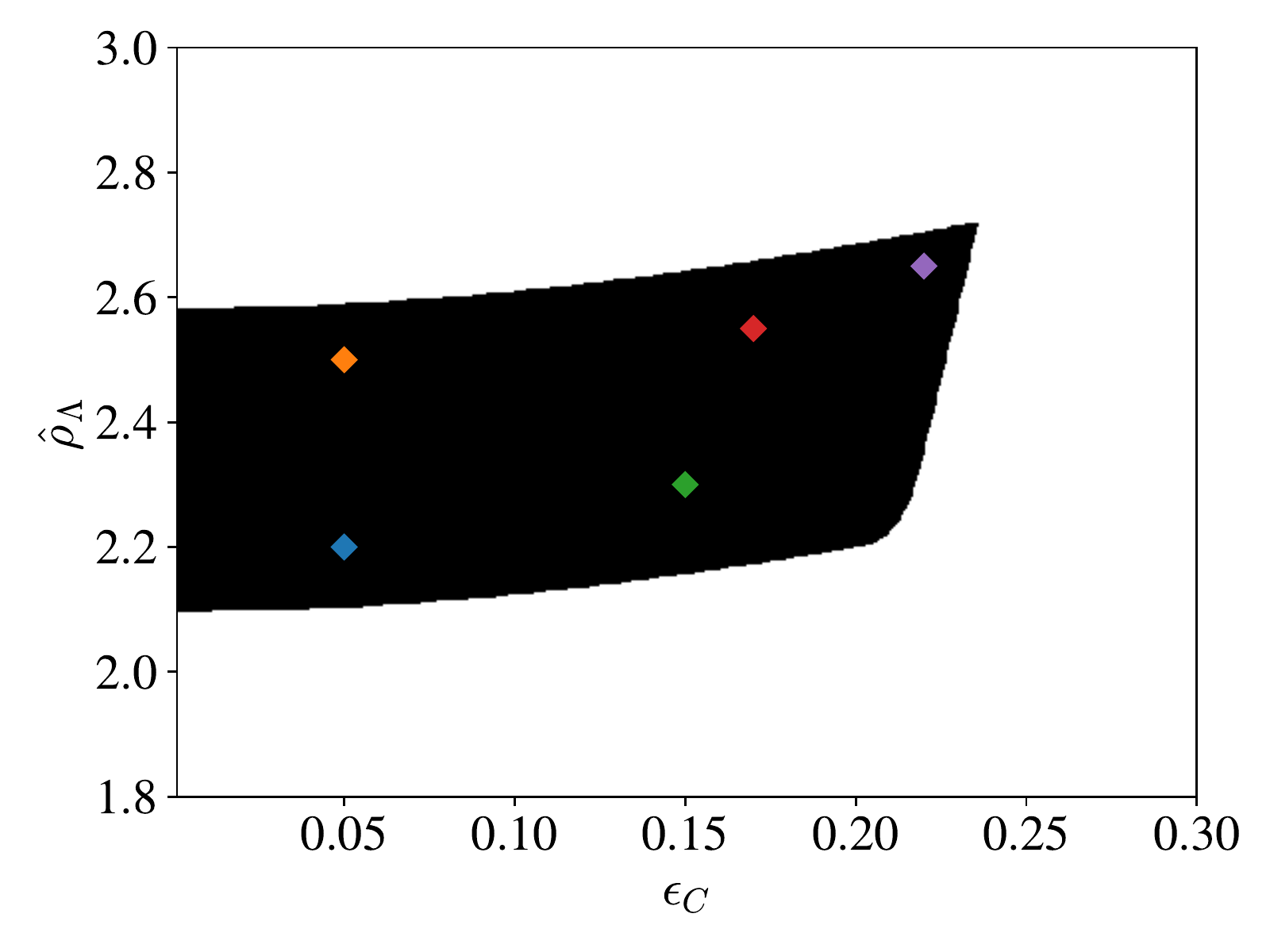}
    \caption{Positions of a selection of parameters in the  region of parameter space where the cosmological tests are satisfied: \crule[tabblue]{8pt}{8pt}~$\epsilon_C=0.05,\hat{\rho}_\Lambda=2.2$; \crule[taborange]{8pt}{8pt}~$\epsilon_C=0.05,\hat{\rho}_\Lambda=2.5$; \crule[tabgreen]{8pt}{8pt}~$\epsilon_C=0.15,\hat{\rho}_\Lambda=2.3$; \crule[tabred]{8pt}{8pt}~$\epsilon_C=0.17,\hat{\rho}_\Lambda=2.55$; \crule[tabpurple]{8pt}{8pt}~$\epsilon_C=0.22,\hat{\rho}_\Lambda=2.65$.}
    \label{fig:color_positions}
\end{figure*}
\begin{figure*}
\centering
   \begin{subfigure}[b]{0.45\textwidth}
        \centering
        \includegraphics[width=\textwidth]{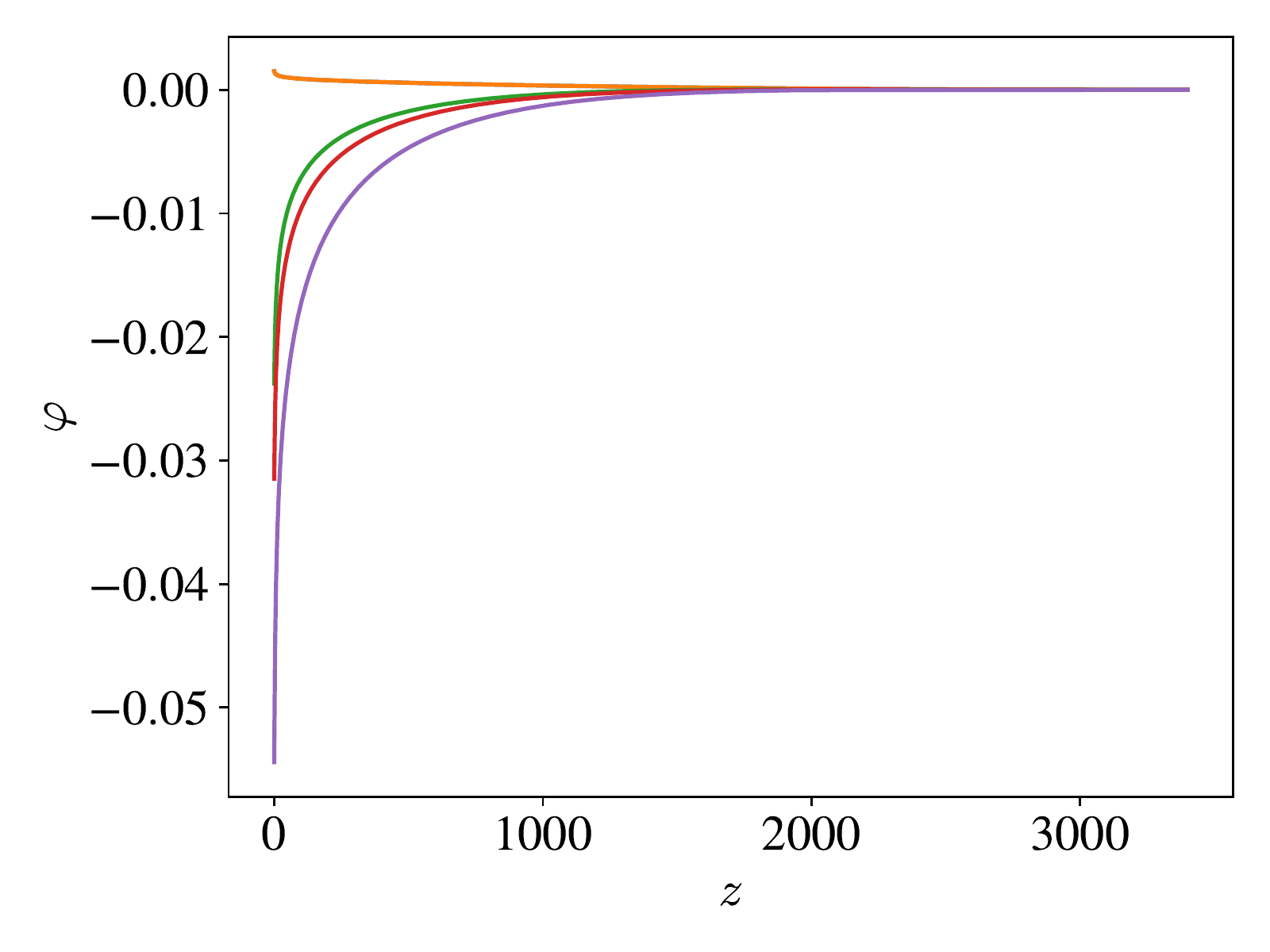}
    \end{subfigure}
   \begin{subfigure}[b]{0.45\textwidth}
        \centering
        \includegraphics[width=\textwidth]{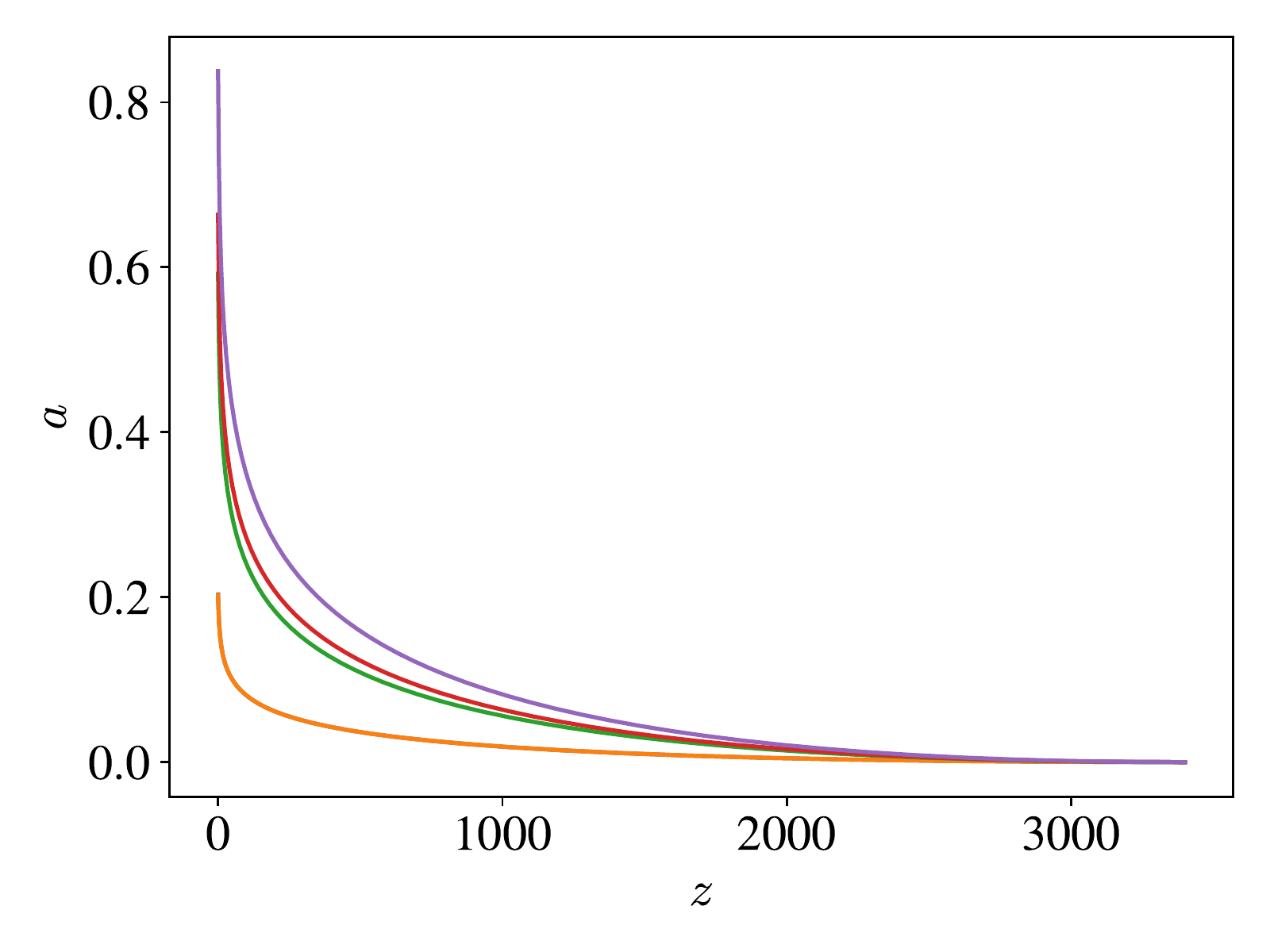}
    \end{subfigure}
\caption{The Field evolutions for $\tauexp=10^{-3}$ and $\epsilon_B=10^{-3}$ as a function of the redshift $z$ defined in the baryon frame. The values of $\epsilon_C$ and $\rhoL$ for each colour are given in \myfigref{fig:color_positions}.}
\label{fig:fields_cosmo}
\end{figure*}
\begin{figure*}
    \centering
    \begin{subfigure}[b]{0.43\textwidth}
        \centering
        \includegraphics[width=\textwidth]{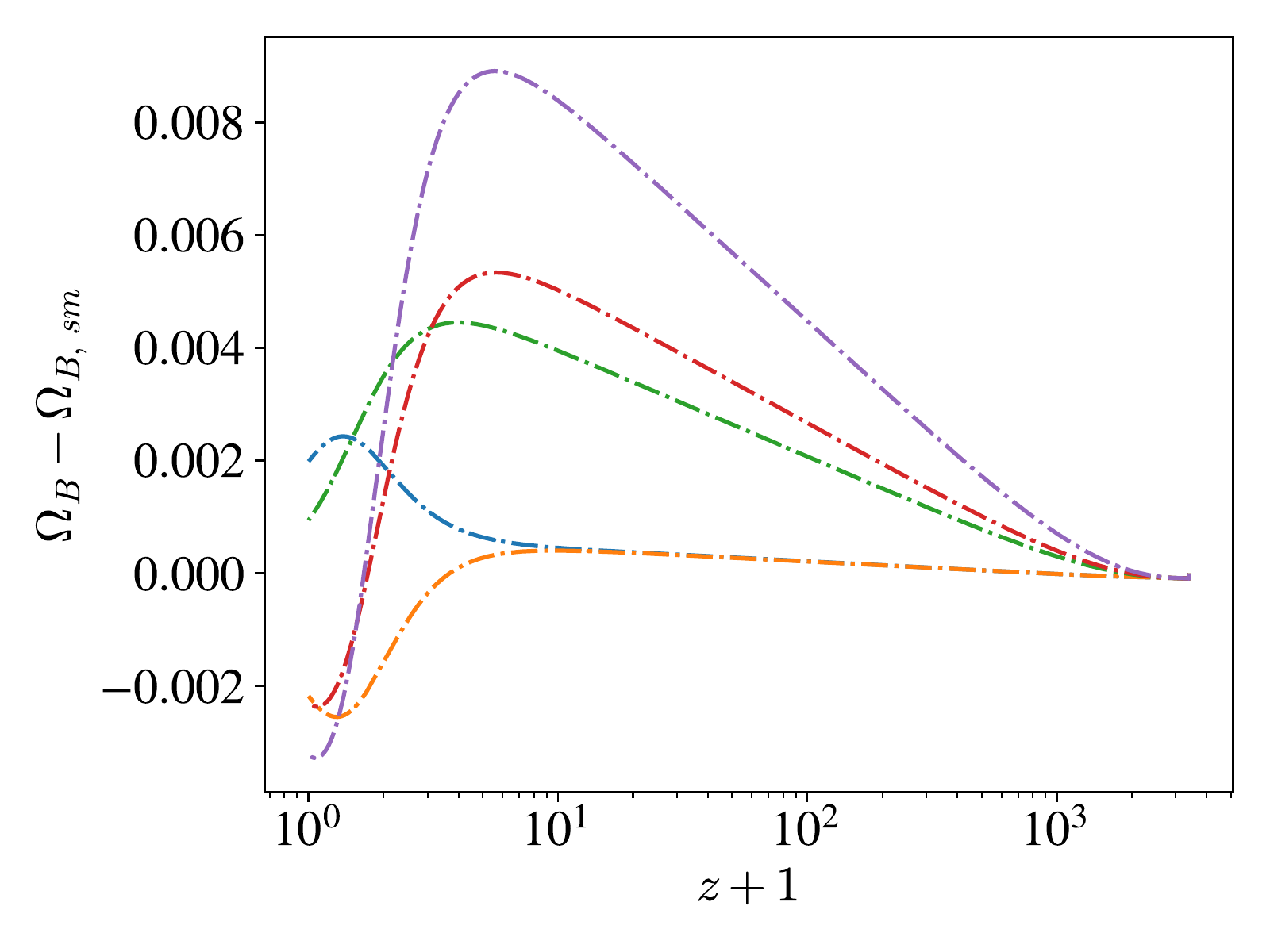}
    \end{subfigure}
    \begin{subfigure}[b]{0.43\textwidth}
        \centering
        \includegraphics[width=\textwidth]{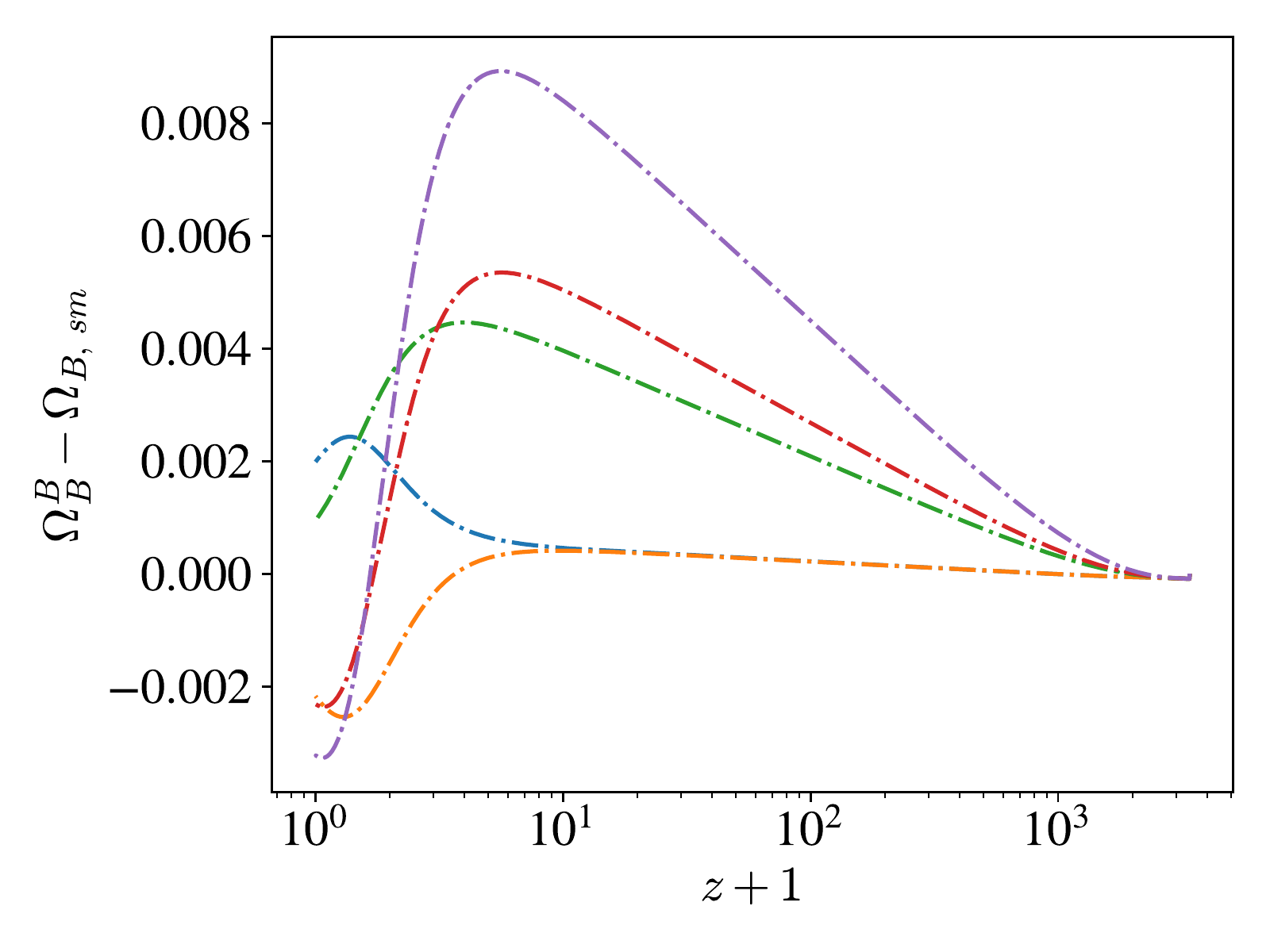}
    \end{subfigure}
    \begin{subfigure}[b]{0.43\textwidth}
        \centering
        \includegraphics[width=\textwidth]{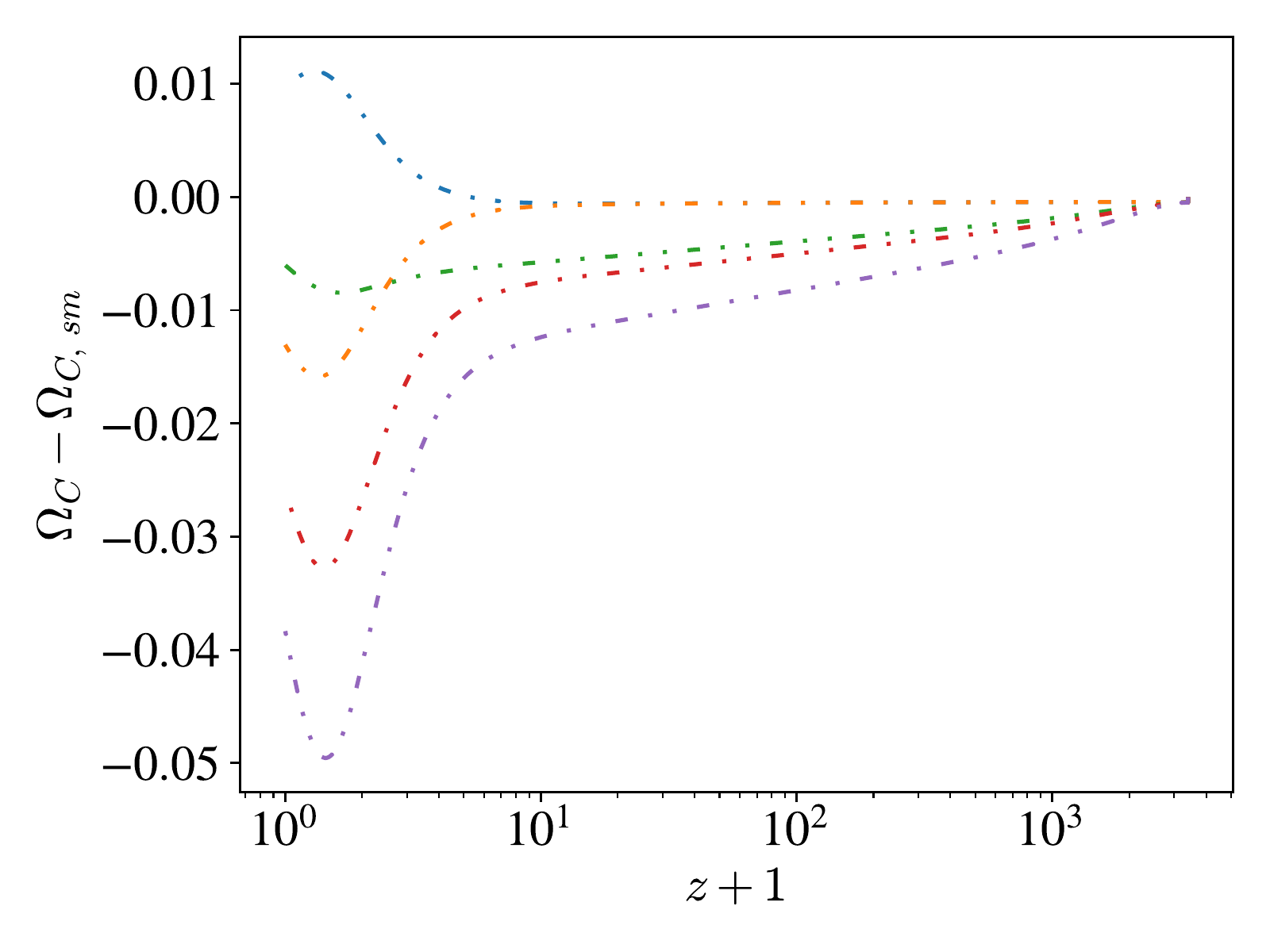}
    \end{subfigure}
    \begin{subfigure}[b]{0.43\textwidth}
        \centering
        \includegraphics[width=\textwidth]{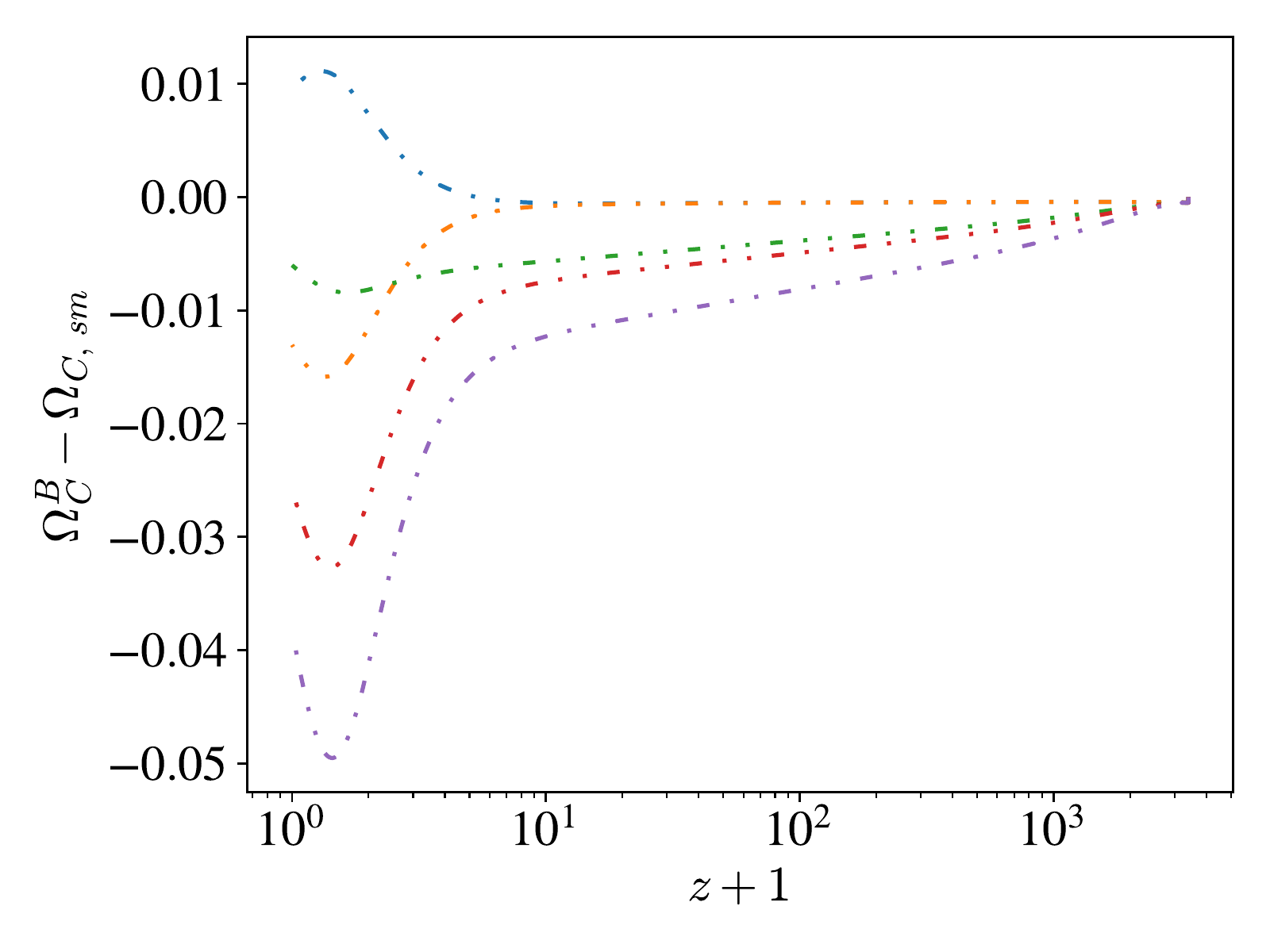}
    \end{subfigure}
    \begin{subfigure}[b]{0.43\textwidth}
        \centering
        \includegraphics[width=\textwidth]{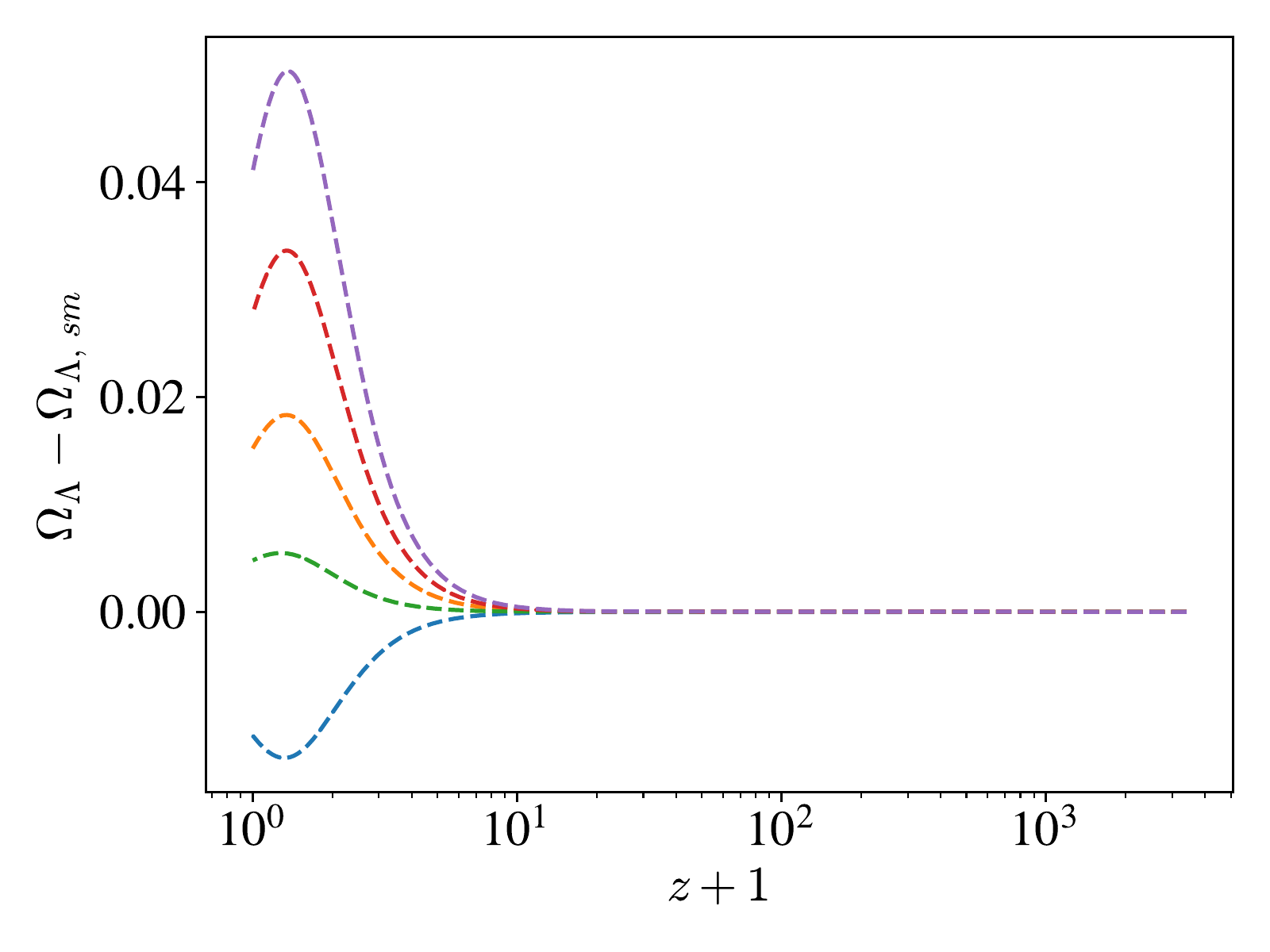}
    \end{subfigure}
    \begin{subfigure}[b]{0.43\textwidth}
        \centering
        \includegraphics[width=\textwidth]{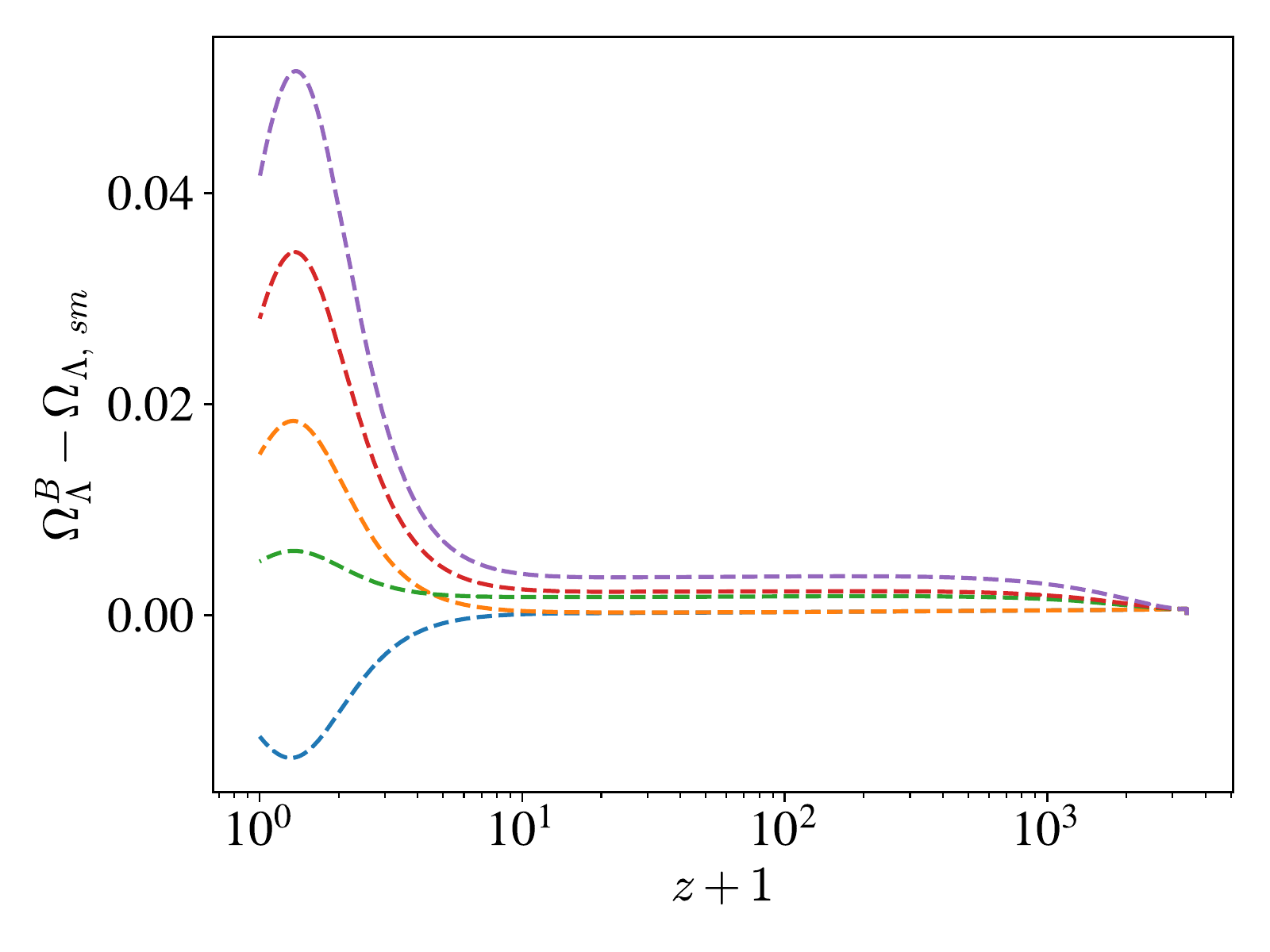}
    \end{subfigure}
    \begin{subfigure}[b]{0.43\textwidth}
        \centering
        \includegraphics[width=\textwidth]{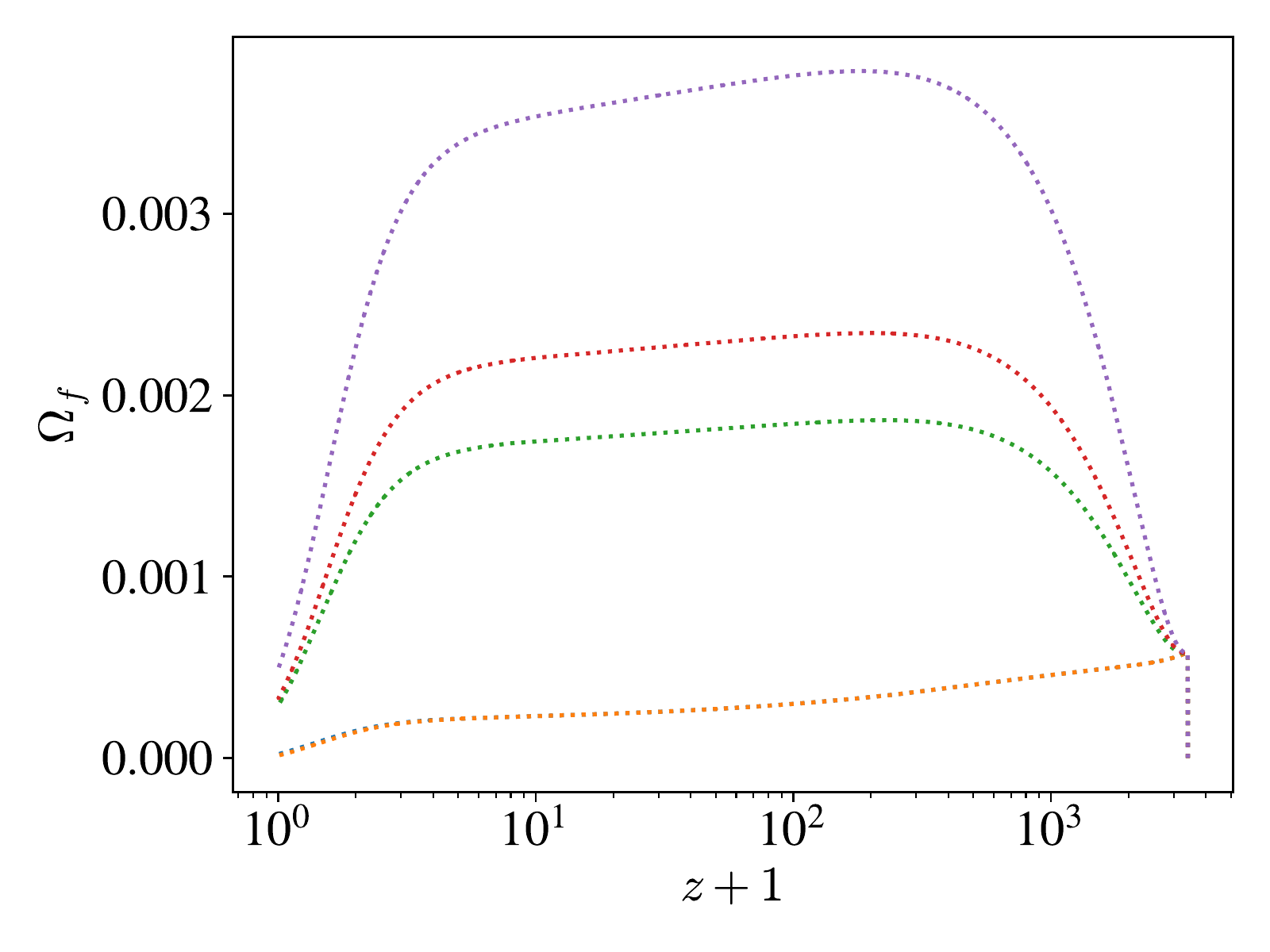}
    \end{subfigure}
    \begin{subfigure}[b]{0.43\textwidth}
        \centering
        \includegraphics[width=\textwidth]{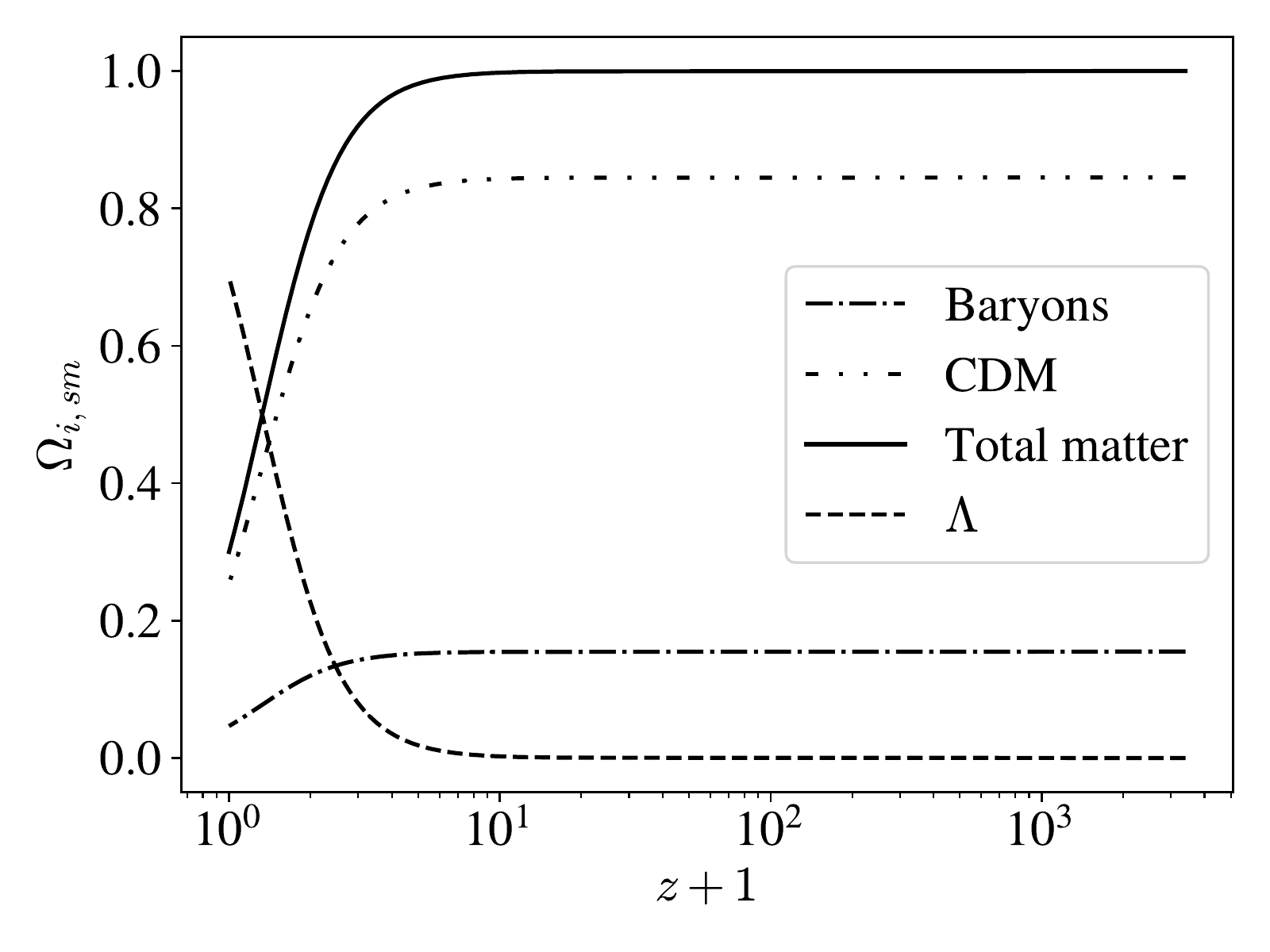}
    \end{subfigure}
    \caption{Evolution of energy contents  of the Universe in the Einstein frame (left) and baryon frame (top three figures on the right) as a function of the redshift of the baryon frame. The $\LCDM$ case  is represented in the bottom right figure. Deviations from $\Lambda$CDM are represented in the other figures for $\tauexp=10^{-3}$, $\epsilon_B=10^{-3}$ and values of $\epsilon_C$ and $\rhoL$ given in \myfigref{fig:color_positions}.}
    \label{fig:cosmo_proportions}
\end{figure*}
\begin{figure*}
    \centering
    \begin{subfigure}[b]{0.45\textwidth}
        \centering
        \includegraphics[width=\textwidth]{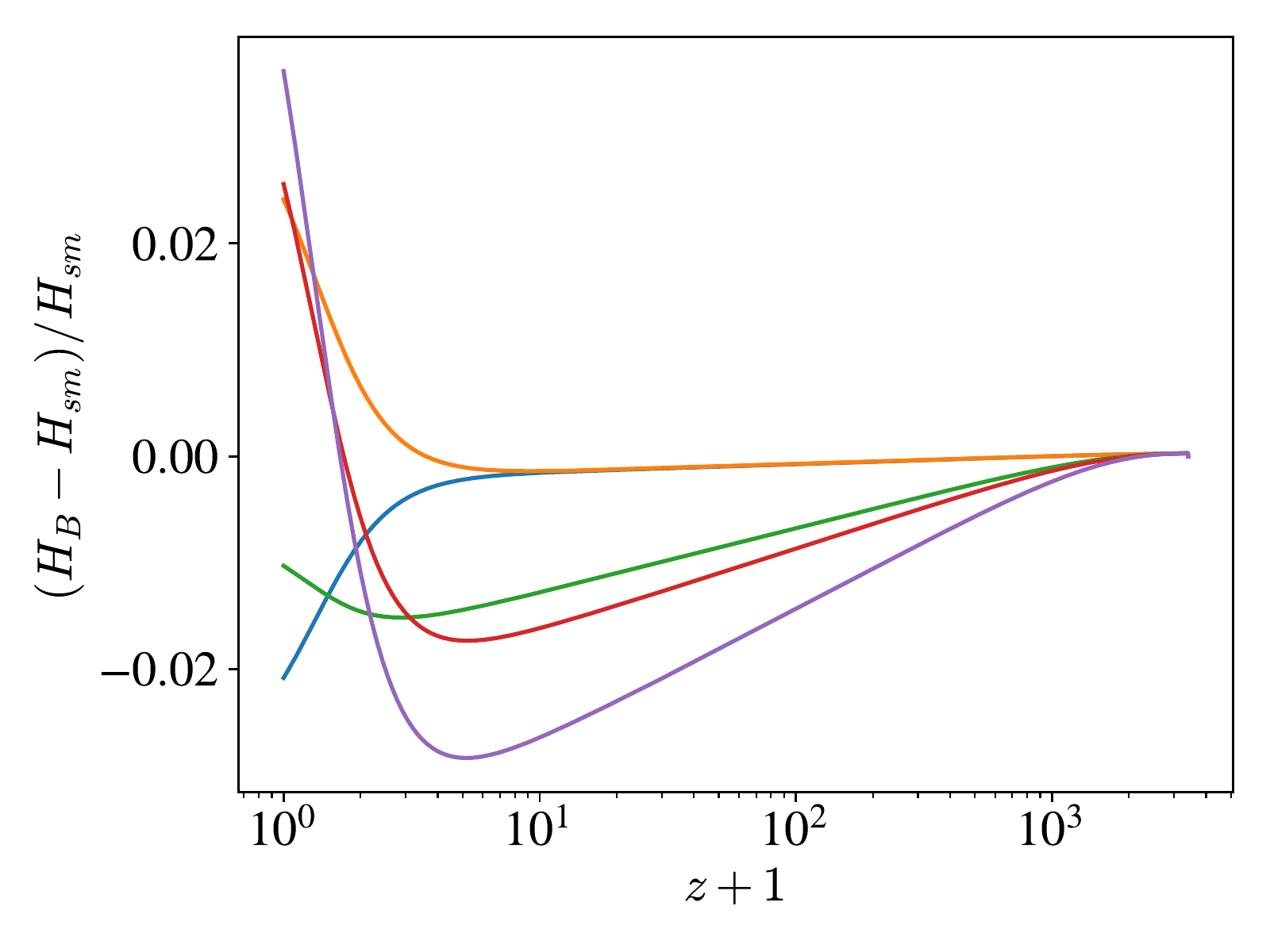}
    \end{subfigure}
    \begin{subfigure}[b]{0.45\textwidth}
        \centering
        \includegraphics[width=\textwidth]{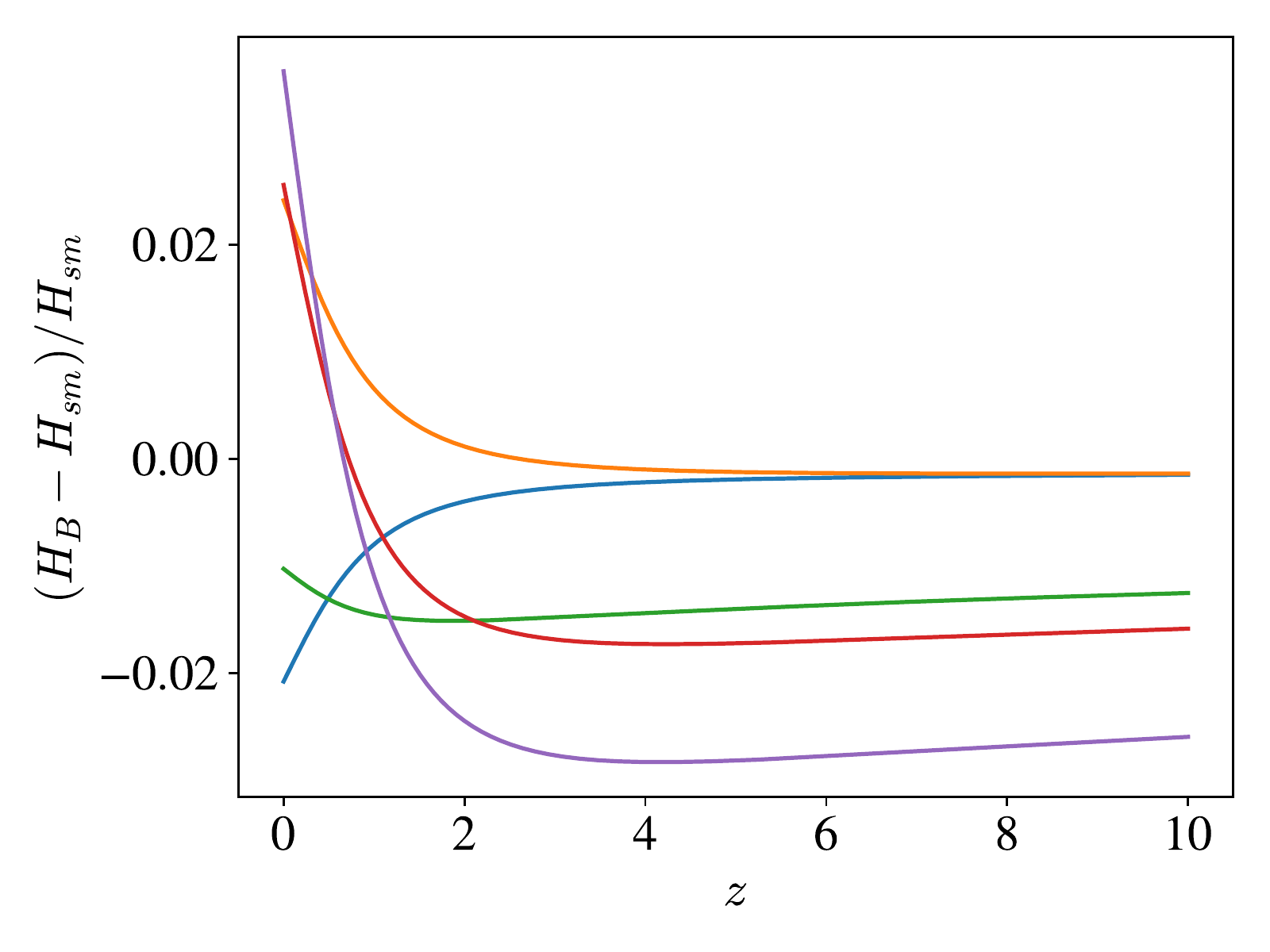}
    \end{subfigure}
    \begin{subfigure}[b]{0.45\textwidth}
        \centering
        \includegraphics[width=\textwidth]{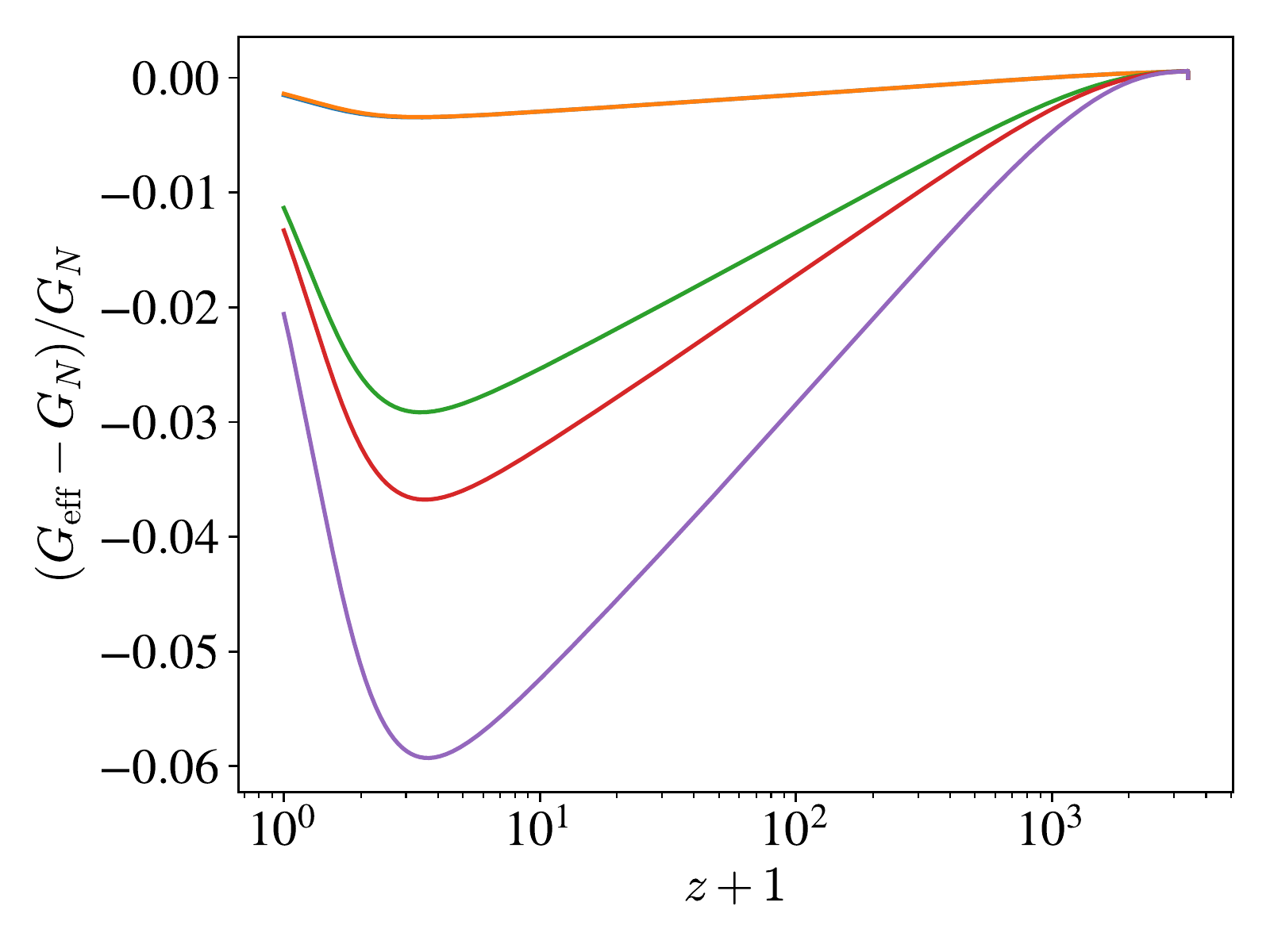}
    \end{subfigure}
    \begin{subfigure}[b]{0.45\textwidth}
        \centering
        \includegraphics[width=\textwidth]{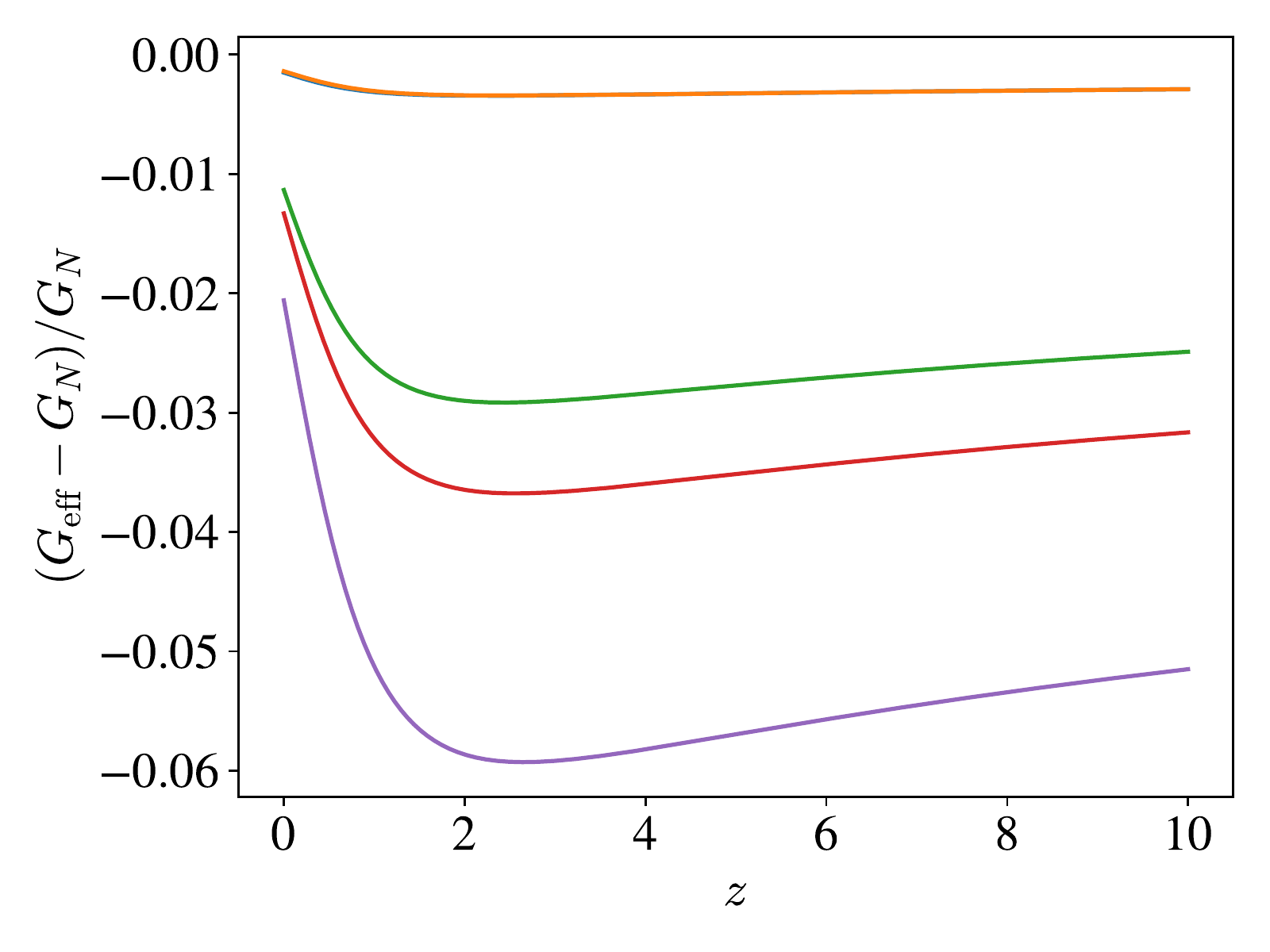}
    \end{subfigure}
    \begin{subfigure}[b]{0.45\textwidth}
        \centering
        \includegraphics[width=\textwidth]{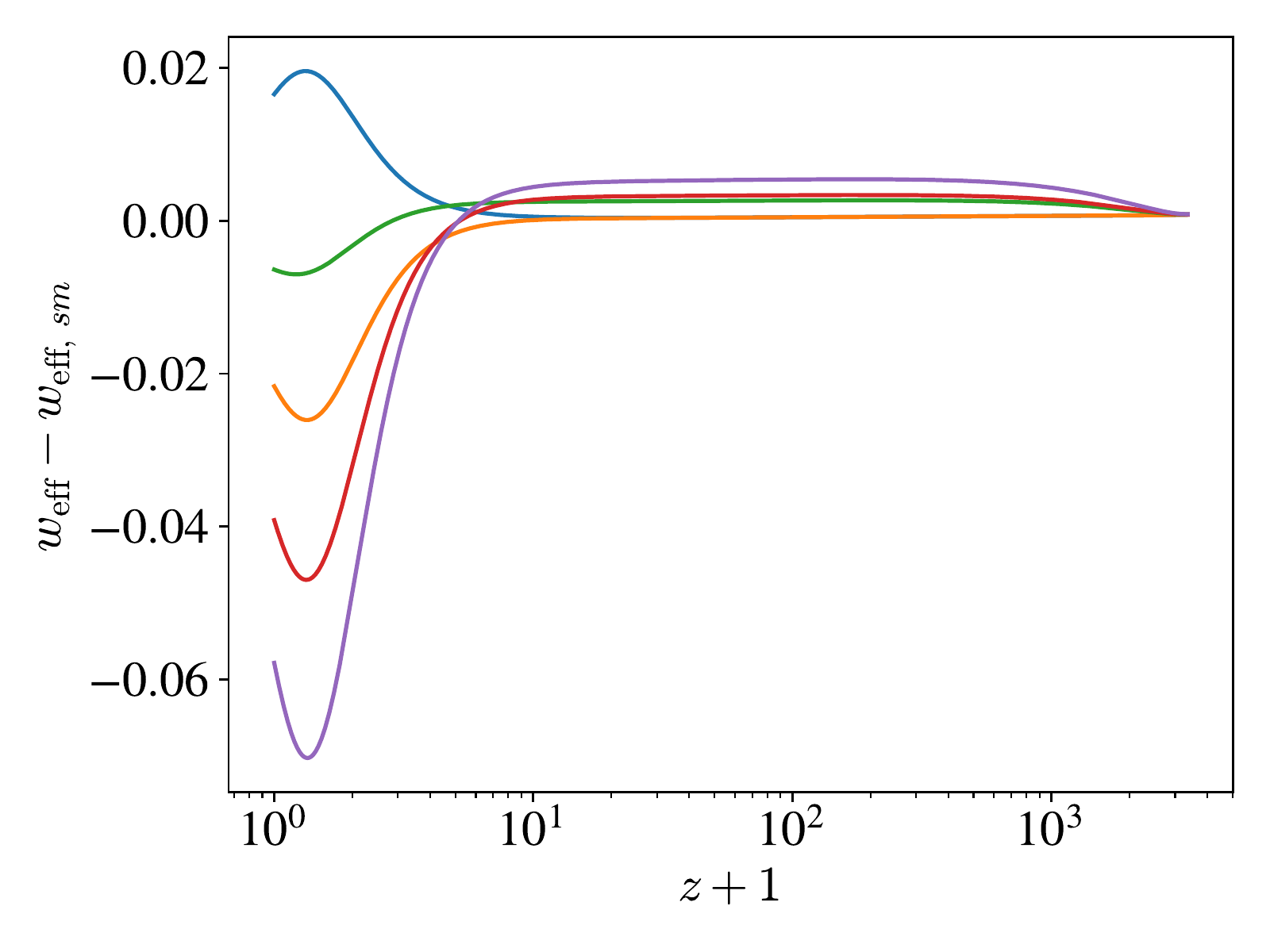}
    \end{subfigure}
    \begin{subfigure}[b]{0.45\textwidth}
        \centering
        \includegraphics[width=\textwidth]{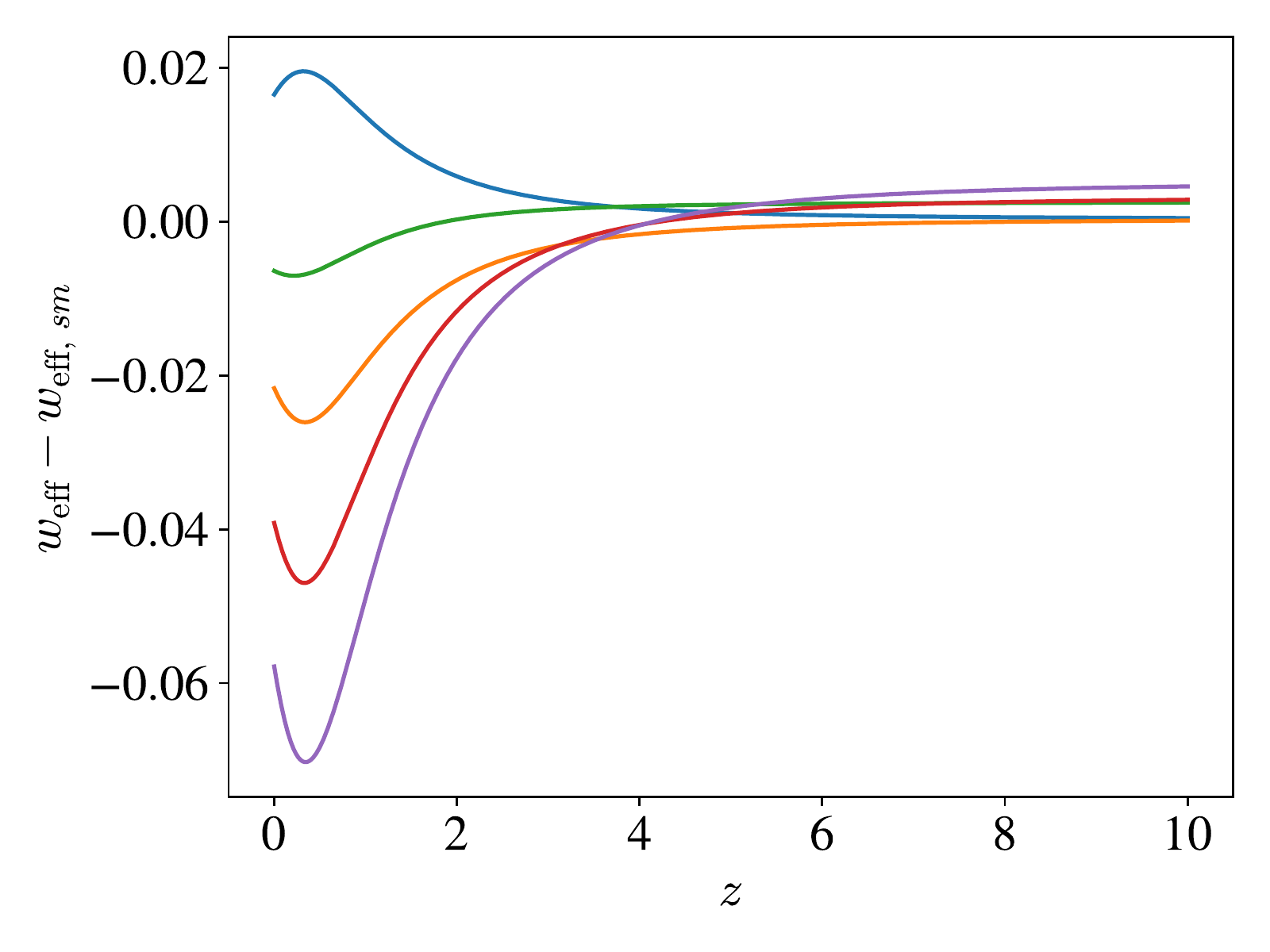}
    \end{subfigure}
    \caption{Evolution of the  cosmological quantities $H_B$, $G_{\rm eff}$ and $w_{\rm eff}$  as a function of the redshift in the baryon frame  for $\tauexp=10^{-3}$ and $\epsilon_B=10^{-3}$.  We zoom on the late evolution on the right. The values of $\epsilon_C$ and $\rhoL$ for each colour are given in \myfigref{fig:color_positions}.}
    \label{fig:cosmo_multiplot}
\end{figure*}
\begin{figure*}
    \centering
    \begin{subfigure}[b]{0.45\textwidth}
        \centering
        \includegraphics[width=\textwidth]{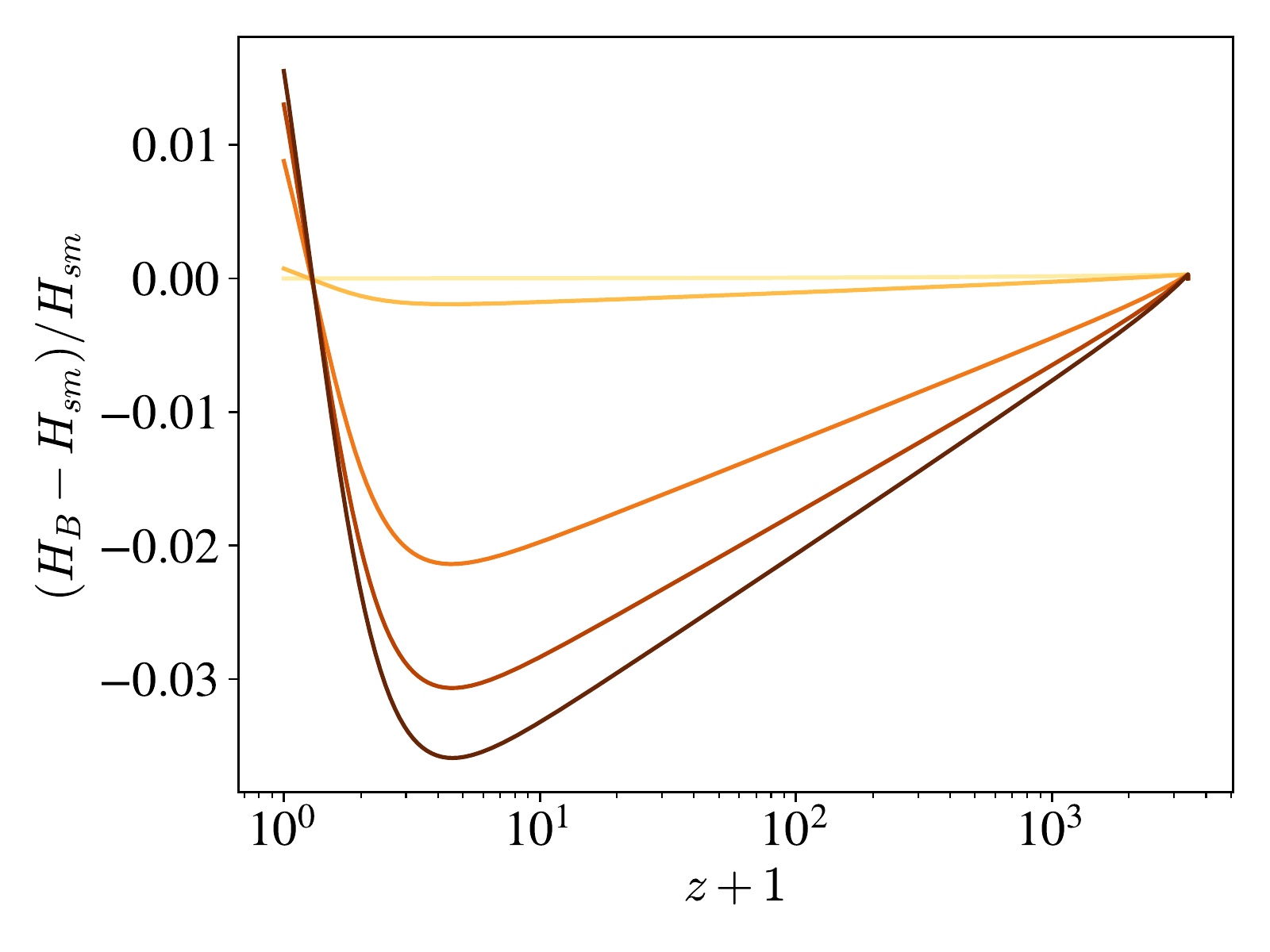}
    \end{subfigure}
    \begin{subfigure}[b]{0.45\textwidth}
        \centering
        \includegraphics[width=\textwidth]{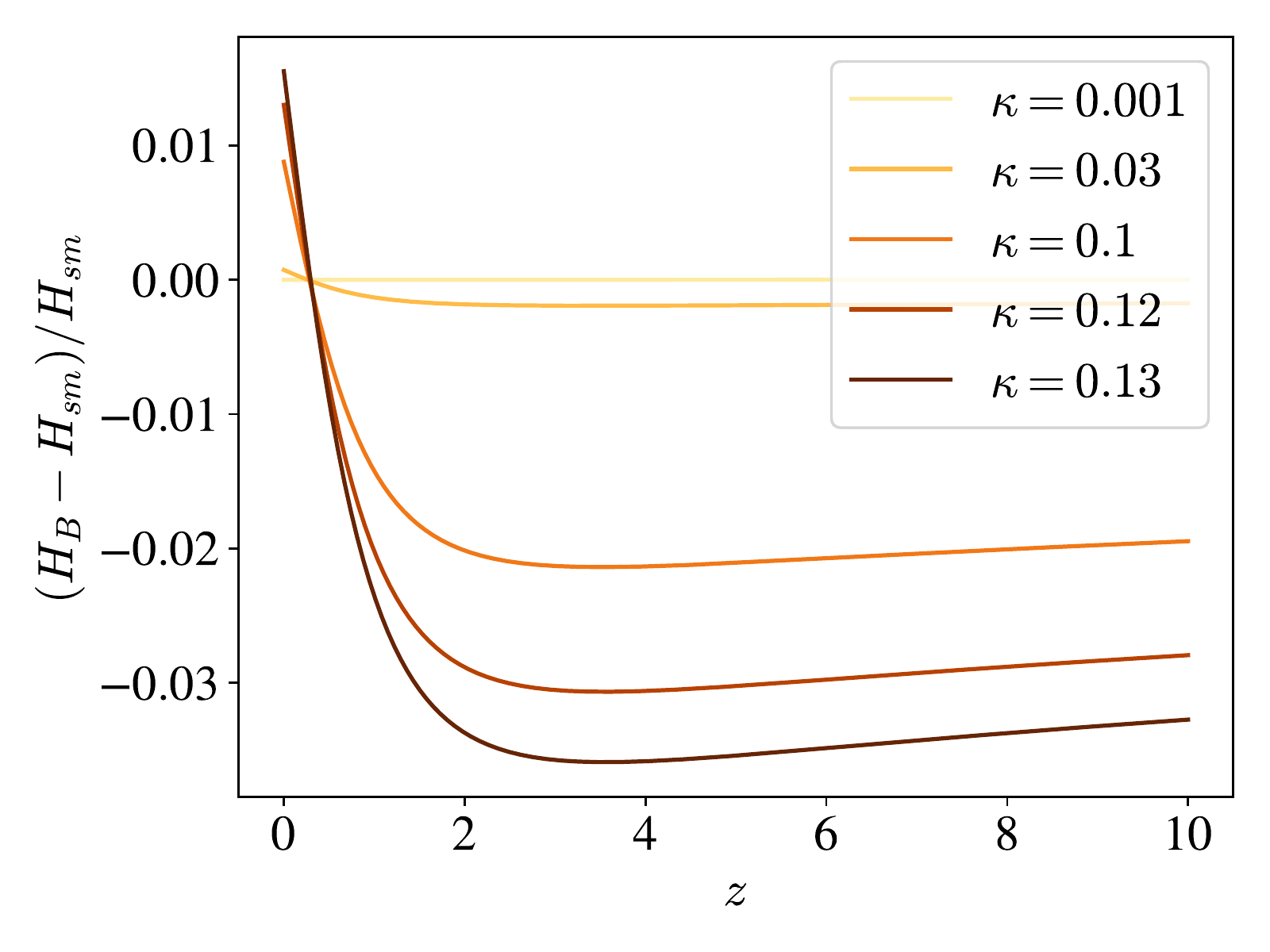}
    \end{subfigure}
    \caption{Deviations of the Hubble rate from $\Lambda$CDM for different values of $\tauexp$, $\epsilon_C=\epsilon_B=10^{-3}$ and $\rhoL=7/3$, as a function of the redshift in the baryon frame.}
    \label{fig:dHH_variations_k}
\end{figure*}

There are four main parameters: $\tauexp$, $\epsilon_B$, $\epsilon_C$ and $\hat{\rho}_\Lambda$. Following our discussion about screening in \mysecref{sec:reevaluating} and our identification of the effective metric in \mysecref{sec:two_fluids}, we will first take  $\tauexp$ and $\epsilon_B$ small. Taking $\tauexp=\epsilon_B=10^{-3}$ turns out to be enough to satisfy solar system constraints. Taking one of them to be one  order of magnitude  higher leads to a violation of  the cosmological constraints when $\epsilon_C$ is large enough.  Numerically we study the cosmological evolution  for a large range of values of $\epsilon_C$ and $\hat{\rho}_\Lambda$. We always start the cosmological evolution at $z_i=3400\approx z_\mathrm{eq}$.

A first picture of the deviations from $\LCDM$ can be obtained by focusing on the cosmological tests (Newton's constant, the equation of state, the Hubble parameter today and in $z\in [0.2,2.5]$ relative to that of the standard model) we defined above and their consequences as viewed in  the $\epsilon-\hat{\rho}_\Lambda$ plane.  This is shown in \myfigref{fig:Observables_k0p001}. We show their values only for the region in which the four of them satisfy the observational constraints. It turns out that {the constraint from BAO is always the strongest}. We illustrate the cosmological evolution using  five sets of parameters in Figs.(\ref{fig:fields_cosmo},\ref{fig:cosmo_multiplot},\ref{fig:cosmo_proportions}). Standard model quantities are computed analytically from the $\LCDM$ solutions in the matter dominated era with $\Omega_{\Lambda,0}/\Omega_{m,0}=7/3$. 

We recover that the smallest deviations to $\LCDM$ are for $\epsilon_C=0$, $\hat{\rho}_\Lambda\simeq 7/3$. We observe that deviations of $H_0$ depend mostly on $\hat{\rho}_\Lambda$ while deviations of $G$ are essentially given by $\epsilon_C$. Additionally, we see deviations which are both positive and negative for $H_0$, the sign depending on the sign of the deviation of $\hat{\rho}_\Lambda$ compared to  $\approx 7/3$. On the other hand, there are only negative deviations of $G$.  As we observe in \myfigref{fig:fields_cosmo}, $\epsilon a$ increases and so $G_{B}$ decreases.

For both $H_0$ and $w_\mathrm{eff}$, the deviations due to $\epsilon_C$ tend to be compensated by deviations of $\hat{\rho}_\Lambda$. This can be understood from the fact that the fields act as a fluid of equation of state $w_f=1$ opposite to that of the cosmological constant $w_\Lambda=-1$. They have opposite effects on the cosmic acceleration. 


Moving on to the dynamical curves, we see in \myfigref{fig:fields_cosmo} the evolution of the fields. The axion $a$ increases with $z$, which can be expected as the source term in its Klein-Gordon equation \eqref{eq:cosmoKGa} has a factor $\epsilon>0$ and initially dominates. On the other hand, $\varphi$ slightly increases at first but eventually decreases much more. This can be expected, as the source term in its Klein-Gordon equation has a factor $\tauexp>0$ which initially dominates, and later the source term is dominated by the axion term proportional to $-\dot{a}^2$. The deviation from the constant and vanishing fields increases with $\epsilon$ as we can expect.

\myfigref{fig:cosmo_proportions} shows that the evolution of the energy content of  the Universe in both the Einstein and baryon frames. As the couplings $\kappa$ and $\epsilon_B$ are small, the difference between the Einstein and the baryon frame quantities is negligible.  In the matter-dominated era we have a slight increase in the field density and a corresponding decrease of the matter density. In the very late Universe close to a vanishing redshift the proportion of the cosmological constant increases and we get the usual values 0.3 and 0.7 for matter and $\Lambda$ with some deviations of the order of 0.01.

As a result we can see that generically the matter contents of the Universe are modified. First of all, the axion and dilaton energy densities evolve from being negligible initially to a long phase in the matter era where they remain nearly constant before dropping to lower values in the last few efoldings of the Universe. This has important consequences as the dynamics of the axio-dilaton system, and in particular the variation of the conformal factors $B_{B,C}$ imply that the matter fractions of the Universe deviate  from their $\Lambda$CDM values. We also notice that the deviation can be negative by a few percent. This is important as this will hamper the growth of structure and entail a compensation of the extra growth due to the attractive scalar forces by the lower amount of matter in the Universe. This will result in a reduced growth of structure in these models.

\myfigref{fig:cosmo_multiplot} shows the evolution for the cosmological quantities of interest.  In $\LCDM$, both $\Delta H_0/H_0$ and $\Delta G/G$ vanish. We observe deviations that gets stronger as the parameters move away from the light regions and into the darker ones of \myfigref{fig:Observables_k0p001}. Both also present a maximal negative deviation around  $z\sim 2$. At smaller redshifts the deviation shrinks. This can be attributed to the effect of the cosmological constant which becomes non-negligible at late times. In the case of the red, orange and purple curves, the deviation crosses $0$ and becomes positive. They have in common that their values for the cosmological constant are stronger than the standard model value: $\hat{\rho}_\Lambda\gtrsim 7/3$. Finally, notice that the evolution of $G_{\rm eff}$ shows a negative deviation from Newton's constant in GR. This has two origins which can be traced to its definition (\ref{Geff}). First of all, as the effective Newtonian constant is normalised with the Planck normalisation for $\rho_0$, the fact that in Fig. \ref{fig:cosmo_proportions} we find that there is less matter in the recent Universe than in $\Lambda$CDM implies that $G_{\rm eff}/G_N $ should be less than unity. Another important effect is that the conformal factors $B_{B,C}$ are less than unity too, implying a reduction of $G_{\rm eff}$ compared to $G_N$. These effects could have been compensated by the extra pull arising from the scalar forces. This is not the case. 

\myfigref{fig:cosmo_multiplot} also shows $w_\mathrm{eff}$. It is defined by generalizing \eqref{eq:weff_def} 
\begin{equation}
     w_\mathrm{eff}(z):=\frac{1}{3}\pqty{\frac{2q_{J}(z)-\Omega_{m,0}}{\Omega_{\Lambda,0}}-1}\qdot
\end{equation}
This can simply be seen as a rescaling of the deceleration parameter $q(z)$. In $\LCDM$ it is easy to see that $w_\mathrm{eff}$ goes from $0$ to $-1$ (and equivalently, $q$ from $0.5$ to $-0.55$). \myfigref{fig:cosmo_multiplot} shows that for parameters consistent with observation the deviations are not drastic.

\subsubsection{Cosmological constraint on \texorpdfstring{$\tauexp$}{TEXT}}
We now concentrate on another scenario where the coupling of the dilaton is relaxed from its solar system bound. This is of interest if the dilaton is screened in the solar system and partially on cosmological scales. The largest dilaton coupling allowed by cosmological data is smaller than the supergravity motivated value $\kappa=1$.

We fix $\epsilon_B=10^{-3}$, and we look for the highest value of  $\tauexp$ such that there are some values of $\epsilon_C$ and $\rhoL$ such that the observables  are within the observational bounds. 
 We use a precision of the order $10^{-3}$ relative to the order of magnitude of the parameters.
We find that the maximal order of magnitude of $\kappa$ is $0.1$. More precisely, if we fix also $\epsilon_C$ and try to find some valid values of $\rhoL$, we find that for $\epsilon_C=0.1$, $\tauexp_{\rm lim}=0.110$; and for $\epsilon_C=0.001$, $\tauexp_{\rm lim}=0.124$.

This is illustrated in \myfigref{fig:dHH_variations_k}. Since BAO is the strongest constraint, we look only at the deviations of the Hubble rate in the BAO interval. We can see that for favourable values of the other parameters ($\epsilon_C=\epsilon_B=10^{-3}$, $\rhoL=7/3$), $\tauexp$ is allowed at least up to $0.12$. But for $\tauexp=0.13$ the BAO constraint is no longer satisfied. This is of course much less than unity and signals that the dilaton must be cosmologically screeened.

\section{Cosmological perturbations}\label{sec:perturbations}
In this section, we study the growth of perturbations for the axio-dilaton models in the matter era in the sub-horizon and quasi-static approximation \cite{Pogosian:2016pwr}. We will obtain the equation governing the evolution of  the density contrast $\delta\rho/\rho$ \cite{Zhao:2008bn}.

\subsection{Cosmological perturbations}

We focus on small perturbations to the background solutions. 
We are interested in structure formation and will evaluate the time evolution of the baryon and CDM density contrasts simultaneously \cite{Huterer:2013xky}. 
We only consider the scalar modes of the perturbation of the metric. Using Newton's gauge in the Einstein frame we have
\begin{equation}
    {g}_{\mu\nu}dx^\mu dx^\nu=R^2(\eta)(-({1+2\Phi_N})d \eta^2+(1-2\Phi_N)\gamma_{i j}d x^i d x^j)
\end{equation}
where $\gamma_{ij}=\delta_{ij}$.
The cosmological perturbations are defined by
\begin{equation}
    \rho_i=\overline{\rho}_i+\delta\rho_i=\overline{\rho}_i(1+\delta_i), \  u^\mu_i=\frac{1}{R}(1+\delta v^0\,,\,\vec v_i)
\end{equation}
for each field. 
We also perturb the fields and the axion-matter coupling:
\begin{equation}
\varphi=\overline{\varphi}+\delta\varphi \qcom
\end{equation}
\begin{equation}
    a=\overline{a}+\delta a\qcom
\end{equation}
\begin{equation}
    \cA=\overline{\cA}+\delta\cA\qdot
\end{equation}
All the symbols with a bar  denote  background values.
The velocity $u^\mu_i$ of the two fluids are given by
\begin{equation}
    u^\mu_i=\frac{1}{R}(1-\Phi_N\,,\,\vec v_i)\qdot
\end{equation}
at linear order. 
The perturbed Newton's law becomes for each species
\begin{equation}\label{eq:CPT_NLi_epscste}
    \partial_\eta \vec v_i+(\cH+\partial_\eta\overline{\chi}_i)\vec v_i=-\frac{\vec \nabla}{R^2}(\Phi_N+\delta\chi_i).
\end{equation}
We recognise the scalar force in the Euler's equation due to the interaction with dark matter and the friction term whose origin is the modified Hubble rate ${\cal H}_i= \cH+\partial_\eta\overline{\chi}_i$ in conformal time and in the Einstein frame.

\begin{figure*}
    \centering
    \begin{subfigure}[b]{0.8\textwidth}
    \includegraphics[width=\textwidth]{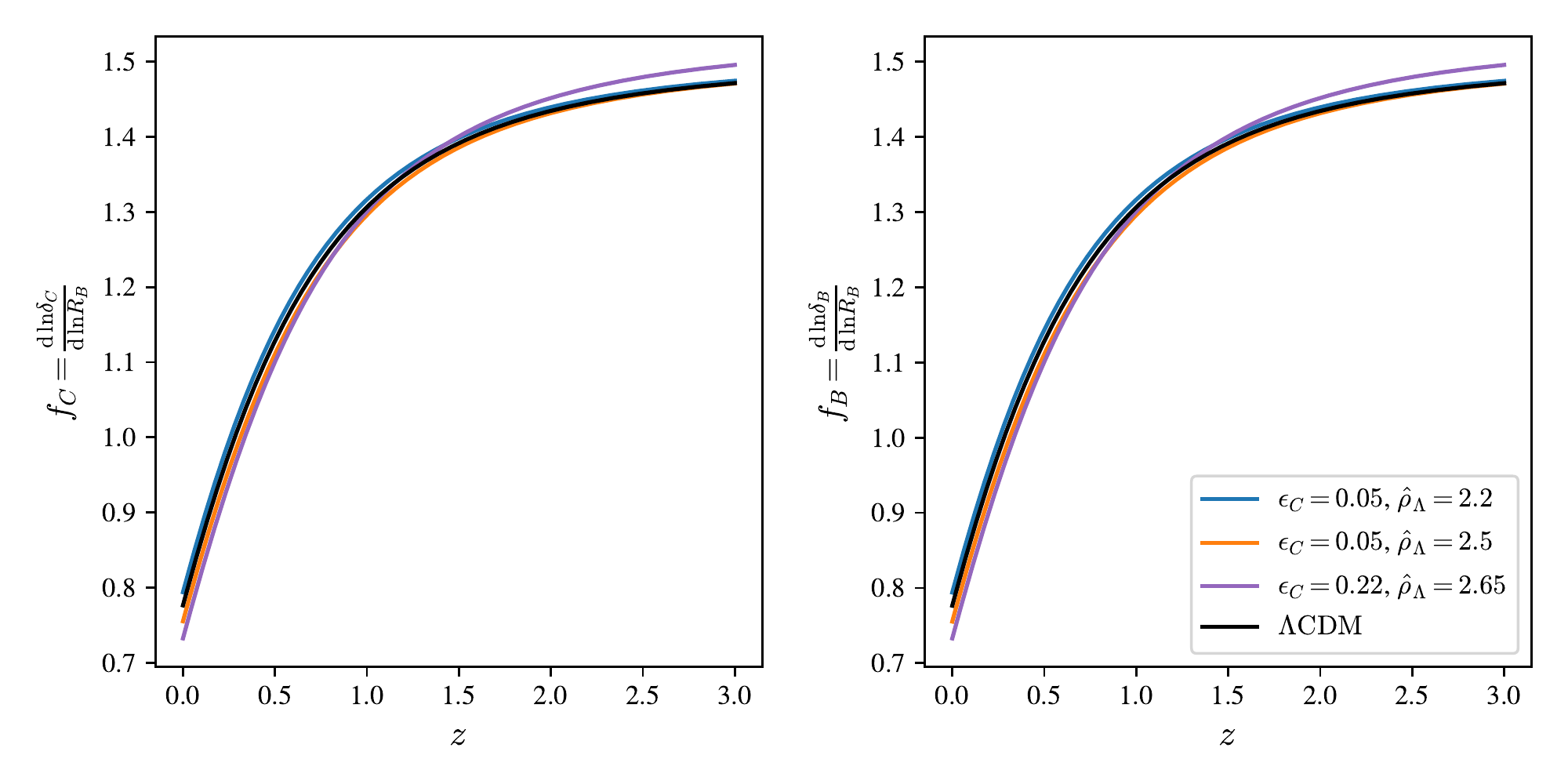}
    \end{subfigure}
    \begin{subfigure}[b]{0.8\textwidth}
    \includegraphics[width=\textwidth]{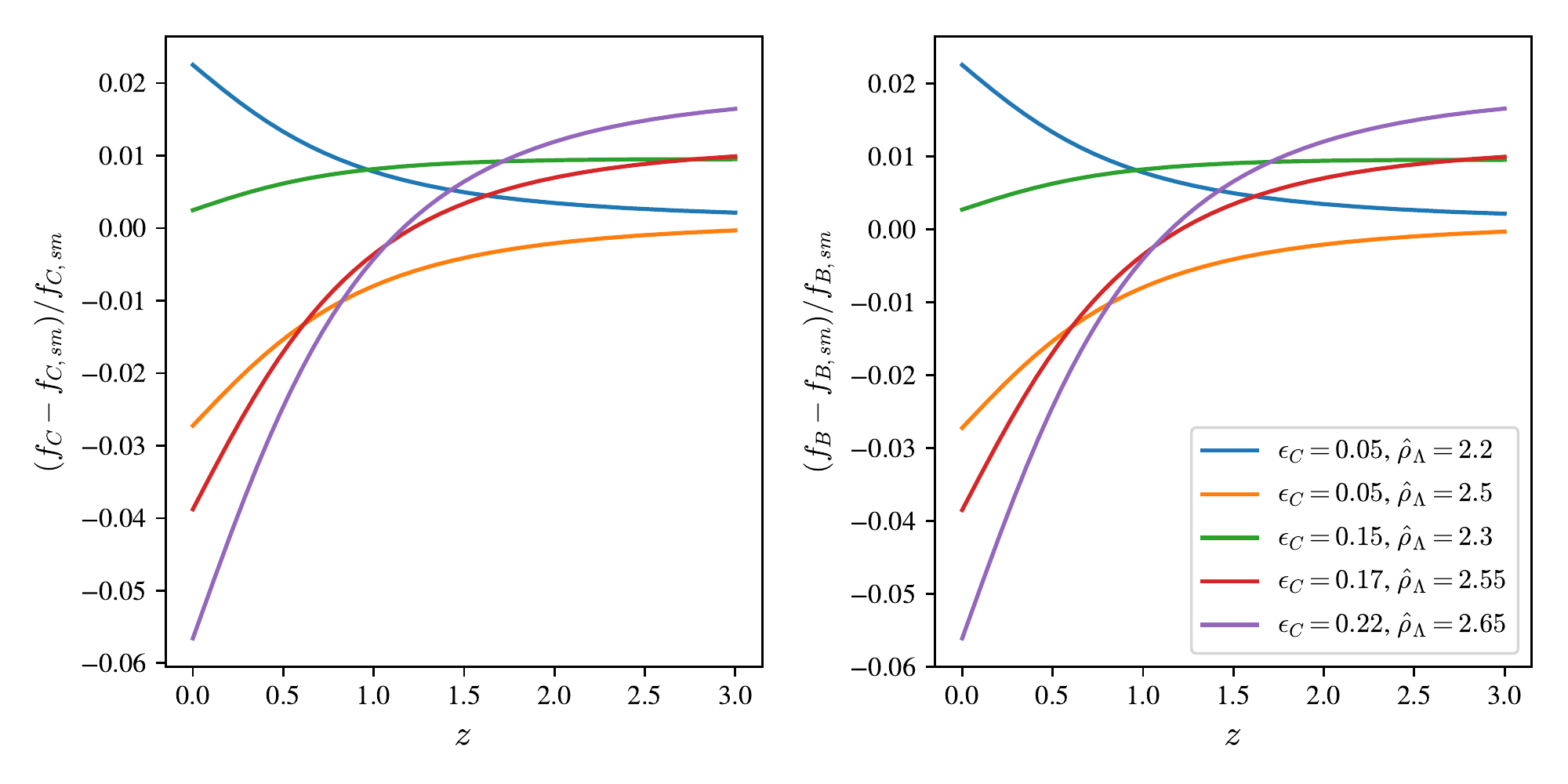}
    \end{subfigure}
    \caption{Growth rate for a few allowed parameters and in $\LCDM$ shown for redshifts in the baryon frame $z\leq 3$. }
    \label{fig:growth_rate}
\end{figure*}
\begin{figure*}
    \centering
        \includegraphics[width=0.8\textwidth]{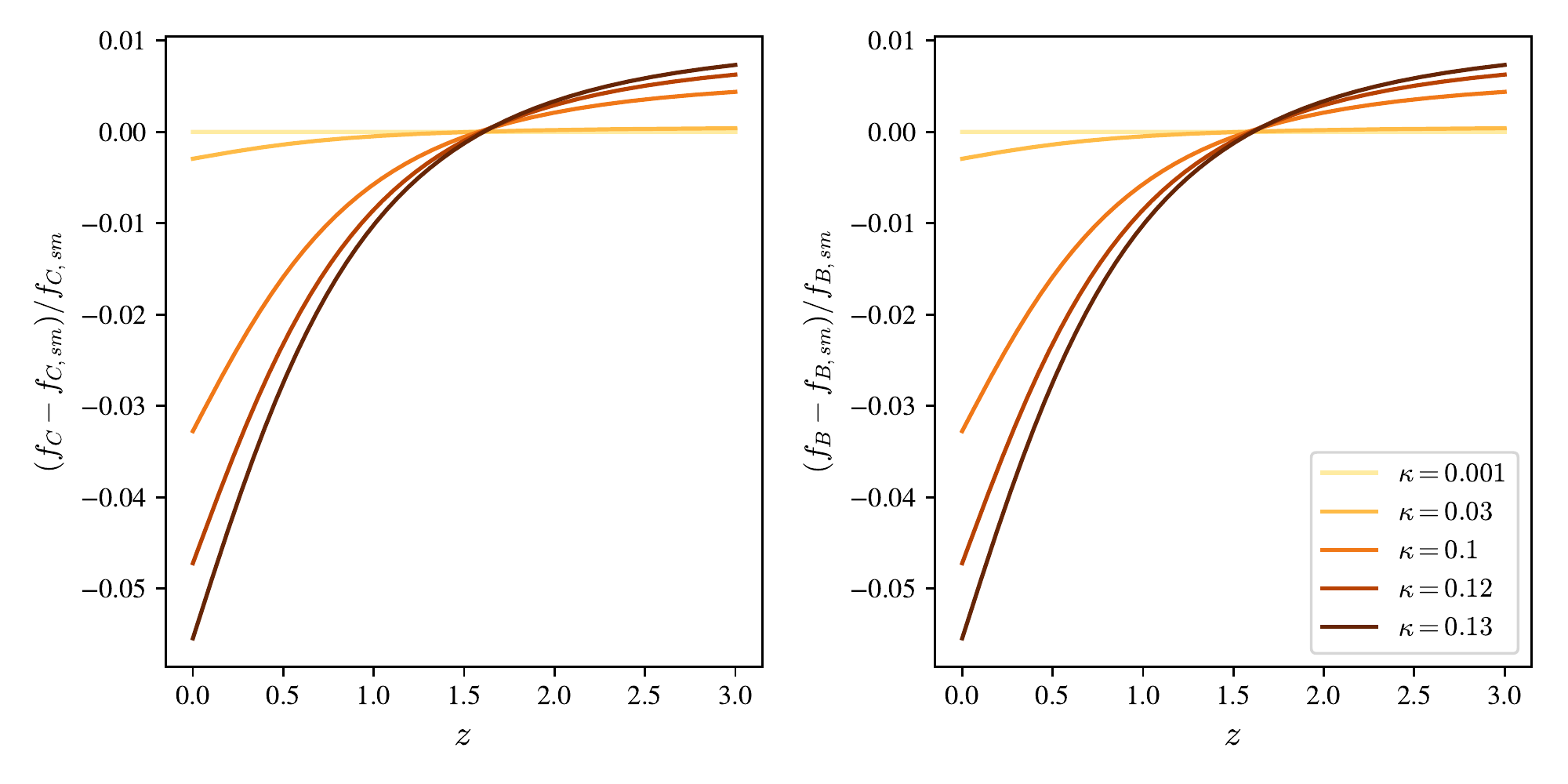}
    \caption{Growth rate at $z\leq 3$ in the baryon frame for different values of $\tauexp$ and $\epsilon_C=\epsilon_B=10^{-3}$, $\rhoL=7/3$.}
    \label{fig:growth_variations_k}
\end{figure*}

\subsection{Density contrast of sub-horizon modes in quasi-static regime}
 We focus on sub-horizon modes such that 
$
k \gg \cH 
$ 
and in the quasi-static regime $\partial_\eta \sim \cH$.
Using the Einstein equation
\begin{equation}
\ricci_{00}-\frac{3}{2\tau^2}(\partial_\eta\tau\partial_\eta\tau+\partial_\eta a\partial_\eta a)-\frac{1}{M_p^2}(T_{00}-\frac{1}{2}T g_{00})=0 \qcom
\end{equation}
and  the curvature perturbation
\begin{equation}
\delta\ricci_{00}=3\cH \Phi'_N+\Delta\Phi_N\approx\Delta\Phi_N\qcom
\end{equation}
which reduces to the Laplacian of Newton's potential as $3\cH \Phi'_N\sim\cH^2 \Phi_N \ll k^2\Phi_N \sim \Delta\Phi_N$, the Poisson equation then becomes
\begin{equation}\label{eq:CPT_Poisson}
    \Delta\Phi_N=\frac{1}{2M_p^2}R^2(\bar \rho_B \delta _B +\bar \rho_C\delta_C)\qdot
\end{equation}
The perturbed Klein-Gordon equation for $\varphi$ is
\begin{equation}\label{eq:CPT_Poisson_phi}
    \Delta\delta\varphi=-\frac{\tauexp}{3M_p^2}R^2(\bar \rho_B \delta _B +\bar \rho_C\delta_C)\qdot
\end{equation}
Its structure is similar to the Poisson equation. 
Similarly we obtain for the axion field
\begin{equation}\label{eq:CPT_Poisson_a}
    \Delta\delta a =-\frac{e^{2\overline{\varphi}}}{3M_p^2}R^2(\bar \rho_B \epsilon_B \delta _B +\bar \rho_C\epsilon_C\delta_C)
\end{equation}
where we have systematically used the sub-horizon and quasi-static approximations. 
The conservation  equation for each species implies that
$
\rho_i= B_i \rho_{i\rm cons}
$
where the conserved density satisfies
\begin{equation}
    \frac{d\rho_{i\rm con}}{d\tau_i}+3h_i\rho_{i \rm con}=0.
\end{equation}
 In the subhorizon limit,  we can identify $\delta_i \simeq \delta_{i\rm cons}$. This is also the density constrast in the baryon frame as the contributions of both $\delta a$ and $\delta \phi$ to the change of frame are negligible in the subhorizon limit. This  implies  that the perturbed conservation equation becomes
\begin{equation}
    \delta'_i=-\vec \nabla. \vec v_i\qdot
\end{equation}
Now we apply $\vec \nabla$ to \myeqref{eq:CPT_NLi_epscste} and obtain
\begin{equation}
-\delta_i''-(\cH+\overline{\chi}_i')\delta_i'=-\frac{\Delta}{R^2}(\Phi_N+\delta\chi_i)\qdot
\end{equation}
Using the Laplacians from \myeqref{eq:CPT_Poisson}, \myeqref{eq:CPT_Poisson_phi}, and \myeqref{eq:CPT_Poisson_a} we finally deduce the growth equation for each species in the subhorizon limit
\begin{eqnarray}
&&\delta_i''+\left(\cH+\overline{\chi}_i'\right)\delta_i'
-\frac{3}{2}\Omega_B\cH^2\left(1+\frac{\tauexp^2+e^{2\overline{\varphi}}\epsilon_B^2}{3}\right)\delta_B \nonumber \\
&&
-\frac{3}{2}\Omega_C\cH^2\left(1+\frac{\tauexp^2+e^{2\overline{\varphi}}\epsilon_C^2}{3}\right)\delta_C
=0 \nonumber \\
\end{eqnarray}
in terms of the matter fraction $\Omega_i$ in the Einstein frame. 
We find that the deviations from the standard model have two origins. First there is the friction term depending on ${\cal H}_i= {\cal H} + \overline{\chi}_i'$ which is specific to each species as the two fluids couple differently to the axion. There is also a modification of Newton's constant for each species on  the two effective couplings 
\be 
G_N^i=(1+2 Q_i^2)G_N
\ee
such that the perturbation equations become 
\begin{eqnarray}
&& \delta_i''+(\cH+\overline{\chi}_i')\delta_i'\nonumber \\ && -\frac{3}{2}\Omega_B\cH^2(1+2Q_B^2)\delta_B -\frac{3}{2}\Omega_C\cH^2(1+2Q_C^2)\delta_C=0\qdot \nonumber \\
\end{eqnarray}
We have defined the effective couplings
\be 
Q_i^2=\frac{\tauexp^2+e^{2\overline{\varphi}}\epsilon^2_i}{6}
\ee
which parameterise the deviations from $\Lambda$CDM. We retrieve that in the absence of the $\epsilon_i$'s,  gravity is modified by a factor $\kappa^2/6$ like in the static regime around a compact object. In the following, we will solve  these equations numerically as a way of investigating the growth of structure for the baryons and CDM.

\subsection{Growth rate}
We now focus on the growth rate for small redshifts \cite{DESI} as defined by
\begin{equation}
    f_i=\frac{\mathrm{d}\,\ln{\delta_i}}{\mathrm{d}\,\ln{R_B}}
\end{equation}
where  the redshift is deduced from  the baryonic scale factor. 
We represent the growth rates in \myfigref{fig:growth_rate}. We choose as initial conditions $\delta_i(z_{ini})\ll 1$ and $\delta_i'(z_{ini})/\delta_i(z_{ini})\sim\cH_{ini}$. We notice that the growth can be either enhanced or disfavoured at small redshifts depending on $\epsilon_C$ and $\rho_\Lambda$. On the other hand, the maximal deviation from $\Lambda$CDM is at most five percent. We have also represented the growth factors when $\kappa$ is varied and both $\epsilon_{B,C}$ are small. Notice that when $\kappa$ is varied, the growth rate at small redshift becomes smaller and smaller. This is also the case when $\kappa$ is fixed and $\epsilon_C$ is increased. As the effective Newton constants should increase the growth thanks to the presence of fifth forces between particles, we conclude that the background evolution and the effective friction have  a drastic effect on the growth of structure. This can be observed in \myfigref{fig:cosmo_proportions} where the matter density in the late Universe decreases compared to $\Lambda$CDM as in \cite{Barros:2018efl} where a similar effect was obtained and used to alleviate the $\sigma_8$ tension. It would certainly be interesting to see if this trend can be also present in the non-linear regime and could have some consequences for the $S_8$ tension where less matter clustering  is observed at late time than inferred in the $\Lambda$CDM scenario \cite{Poulin:2022sgp} from the Planck data \cite{Planck:2018vyg}. The analysis of the $S_8$ tension in this scenario  is left for future work. 

Finally let us remark that we have not taken into account an important effect which would result from both the fact that the effective equation of state $w_{\rm eff}$ (see Fig. \ref{fig:cosmo_multiplot}) is not strictly equal to zero deep into the matter era for $z\gtrsim 2$ and that the perturbations do not behave like in the Einstein-de Sitter Universe with $f\ne 1$ (see Fig. \ref{fig:growth_rate}). This  would imply that the Newtonian potential $\Phi_N$ in (\ref{eq:CPT_Poisson}) is not strictly constant in the matter era. This could lead to a large Integrated Sachs Wolfe effect (ISW) which is tightly constrained and could appear in the galaxy counts versus the CMB (Cosmic Microwave Background) cross-correlations \cite{Crittenden:1995ak,Khosravi:2015boa,Benevento:2018hdb}. This will certainly restrict the available parameter space and constrain the possible deviations of the growth factor from $\Lambda$CDM. A detailed study of this effect is left for future work. 

\section{Conclusion}
\addcontentsline{toc}{section}{Conclusion}
The axio-dilaton model has a clear origin in string theory.  We have focused on  the screening mechanism introduced in \cite{Burgess_2022} for a constant coupling of the axion to matter and explicitly shown how it can only be effective when the field values at infinity are  tuned to specific values. These values should be determined cosmologically.  We have studied the background cosmology of these models and shown that the cosmological values do not correspond to the tuned values generically. 

In the absence of explicit screening for the axio-dilaton system which may require to introduce non-linear couplings to matter and/or new fields \cite{Burgess:2021obw} whose dynamics would drive the couplings of the dilaton and the axion to small values in the solar system and larger values cosmologically, we have employed a simple alternative and considered two scenarios. In the first one, the coupling of the dilaton and the axion to baryons is taken to satisfy the solar system constraints and remain identical to these values on large scales. Only the coupling to cold dark matter is allowed to take much larger values. In this case, we find that the coupling to cold dark matter must be bounded. It turns out that the constraints from Baryonic Acoustic Oscillations (BAO) at small redshift are the tightest and  the present day Hubble rate does not deviate from the Planck normalised one by more than three percent. This is not enough to account for the $H_0$ tension, which lies at the ten percent level. Similarly, the growth of structure is affected at the five percent level. Interestingly, in these models growth is not always enhanced and effectively a decrease in the growth rate is observed for a large part of the parameter space of the model. This follows from the fact that the growth increase due to the scalar forces is compensated by the decrease of the matter density. This could have some relevance to the $\sigma_8$ tension. We hope to come back to this suggestion in the near future.  We also consider the case where the axion does not couple significantly to matter and the dilaton couples with a strength $\kappa$ reduced from the string theory motivated example. We find that $\kappa$ cannot be allowed values of order unity and must be bounded around $0.1$. This entails that the dilaton must not only be screened locally in the solar system, but also cosmologically.

Of course, our examples can be modified and the resulting physics very different. For instance, the couplings to matter of both the axion and the dilaton could become non-linear and therefore lead to screening mechanisms akin to the ones of the symmetron model for instance. Another possibility would be that other fields could relax to values whereby the couplings of the dilaton and the axion would become very small in the solar system and small  cosmologically. The construction of these models is left for future work.

Phenomenologically, the scenarios we have introduced fall within the category of late time dark energy models where 
the evolution of the fields at small redshift would modify both the background cosmology and the growth of structure. As expected, we find that the BAO bound entails a tight constraint on both the possible deviations of the present Hubble rate from its Planck value and the growth factor from its $\Lambda$CDM counterpart. We notice that the allowed deviations of the growth factor could reach a few percent and therefore may become detectable by future large scale surveys. The detailed study of the phenomenology of these models is left for future work. 
\vskip 1 cm 
\noindent{\bf{Acknowledgments:}} 
We would like to thank E. Lindner and L. Pogosian for fruitful comments. We would also like to thank A. T. Ortiz for running some of the long numerical computations.
A.O. would like to thank IPhT for funding during his internship, as well as the physics department of the Ecole Normale Sup\'erieure for its scholarship.

\vskip 1 cm 
\appendix

\section{Validity of the flat background approximation}\label{app}
We consider here the conditions for the flat background approximation used in \mysecref{sec:sourceandscreening} to hold. We  assume the exterior solution of \mysecref{sec:ext_solution} and find in which regime the terms neglected by the approximation are indeed negligible. The metric of \mysecref{sec:sourceandscreening} can be written:
\begin{equation}
    ds^2=-({1+2\Phi})dt^2+(1-2\Psi)(d r^2+r^2d\Omega^2)\qdot
\end{equation}
The Christoffel symbols, the  Ricci and Einstein tensors for this metric can be found in e.g. \cite{carroll} §7. The $(tt)$ and $(rr)$ equations from \myeqref{eq:Einstein} are
\begin{equation}
    \Delta \Phi=\frac{\rho}{2M_p^2}\qcom
\end{equation}
\begin{equation}
    (\Delta-\partial_r^2) ({\Phi-\Psi})=\frac{3}{4}\frac{{(\tau')^2+(a')^2}}{\tau^2}\qcom
\end{equation}
where $\Delta$ is the  Laplacian. The exterior solution gives
\begin{equation}
    \frac{{(\tau')^2+(a')^2}}{\tau^2}=\frac{\gamma^2\beta^2}{r^4}\qdot
\end{equation}
Integration gives us then
\begin{equation}
    \Phi=-\frac{GM}{r}\qquad,\qquad\Psi-\Phi= \frac{3}{16}\frac{\gamma^2\beta^2}{r^2}\qdot
\end{equation}
The deviation from the flat metric are therefore negligible for
\begin{equation}
    \abs{\Psi}\ll 1 \qquad,\qquad \Phi\simeq \Psi\qdot
\end{equation}
This is verified in the regime
\begin{equation}
    r\gg GM \qquad,\qquad r\gg \abs{\gamma\beta}\qdot
\end{equation}

\bibliographystyle{apsrev4-1}
\bibliography{refs}

\end{document}